\documentclass{emulateapj}

\newcommand{\dasec}{\hbox{$.\!\!^{\prime\prime}$}}     
\newcommand{\damin}{\hbox{$.\!\!^{\prime}$}}           
\newcommand{\sdeg}{$^{\circ}$}                         
\newcommand{\ddeg}{\hbox{$.\!\!^\circ$}}               

\newcommand{\rmeff}{$RM_{eff}$}                       
\newcommand{\dt}{$\Delta\theta$ }                      
\newcommand{\hii}{H\ {\small II}}                      

\begin{document}


\title{A Wide-area VLA Continuum Survey near the Galactic Center at 6 and 20 cm Wavelengths}
\shorttitle{6 and 20 cm Survey of Galactic Center}

\shortauthors{Law et al.}
\author{C. J. Law\altaffilmark{1,2}}
\author{F. Yusef-Zadeh\altaffilmark{1}}
\author{W. D. Cotton\altaffilmark{3}}

\altaffiltext{1}{Department of Physics and Astronomy, Northwestern University, Evanston, Illinois 60208, USA; claw@science.uva.nl}
\altaffiltext{2}{Astronomical Institute ``Anton Pannekoek'', University of Amsterdam, Kruislaan 403, 1098 SJ Amsterdam, The Netherlands}
\altaffiltext{3}{National Radio Astronomy Observatory, Charlottesville, VA 22903, USA}

\begin{abstract}
We describe the results of a mJy-sensitivity, VLA survey of roughly 1 square degree near the Galactic center at 6 and 20 cm.  Catalogs of compact and filamentary structures are given and compared to previous surveys of the region.  Eight of the unusual, nonthermal radio filaments are detected in 6 cm polarized emission;  three of these are the first such detections, confirming their nonthermal nature.  This survey found emission from a filament at (l,b)=(359.1,0.75), or a projected distance from Sgr A* of 200 pc, greatly extending the latitude range observed with such features.  There is also new evidence for spatial gradients in the 6/20 cm spectral indices of some filaments and we discuss models for these gradients.  In studying compact sources, the combination of spectral index and polarization information allows us to identify pulsar candidates and compact \hii\ regions in the survey.  There is also some evidence that the flux measurements of compact sources may be affected by electron scattering from the interstellar medium in the central few hundred parsecs of the Galaxy.  
\end{abstract}


\section{Introduction}
The Galactic center (GC), at a distance of roughly 8 kpc \citep{r93}, offers us a unique view of the complex and dynamic world of galactic nuclei.  The central 2$\ddeg$5 of the Galaxy includes 5--10\% of Galaxy's molecular gas and star formation activity \citep{m96}, so it is populated with rare objects like nonthermal radio filaments \citep[NRFs;][]{y84}, compact objects \citep{mu05}, compact \hii\ regions \citep{g93}, and supernova remnants \citep{br03}.  However, as the nearest Galactic nucleus, the GC provides a unique opportunity to understand the unusual phenomena found in the centers of other galaxies.

Interferometric radio observations are especially useful in studying the GC region, since they are less hampered by confusion.  Recent technological changes, such as new imaging algorithms and receivers, have enabled more sensitive, systematic study over a wider range of wavelengths \citep{g94,br03,n04}.  In Yusef-Zadeh et al. (2004, hereafter ``Paper I''), we presented a comprehensive, 20 cm, imaging survey of the GC region.  That effort applied the latest imaging algorithms to archived and new VLA data, creating high-fidelity images of the GC region.  However, that work was limited to one frequency and did not discuss polarization, so we were motivated to expand on Paper I with new, 6 and 20 cm polarized continuum surveys of a roughly 1 degree square region, just north of the GC.  Since radio continuum studies at cm wavelengths are sensitive to both thermal and nonthermal processes, a spectral index study can measure their relative contributions and constrain basic parameters, such as supernova rates and star-formation activity \citep{h07}.

One focus of this work is on the unusual NRFs, the long (tens of parsecs), polarized filamentary structures seen at radio wavelengths \citep{y84}.  These filaments are observed only in the GC region, but are found there in abundance, with about 15 known and several dozen candidates in the central two degrees of the Galaxy \citep[][Paper I]{n04}.  The nonthermal spectra and linearly-polarized radio emission from the NRFs indicate that they emit synchrotron radiation.  However, the manner in which the magnetic field in NRFs is amplified and how their electrons are accelerated is still debated \citep[e.g.,][]{h88,b88,s94,y03,y05}.  Several observations have measured the radio and x-ray spectral indices of NRFs \citep[e.g.,][]{l99,l00,l03}, but there has not been much systematic study.  Thus, a goal of this survey work is to find new examples of NRFs and make more robust measurements of their spectral indices, to help better constrain theoretical models.

This work also studies the compact radio sources in the GC region, which are diverse and probe a variety of physical processes.  One application is in the study of the electron population in the GC region.  The high electron density in the GC strongly disperses time-variable radio emission, frustrating efforts to detect radio pulsars in the central few hundred parsecs \citep{c04}.  Surveys of compact sources in the GC can study this by finding evidence of scatter-broadening from sources in or beyond the GC region \citep{l98}.  Also, sensitive surveys of the spectral index of compact sources can be used to search for candidate pulsars that may guide future searches for pulsations \citep[e.g.,][]{n04,l98b}.

This paper describes our efforts using the VLA to survey a large part of the GC region to study its compact and filamentary emission.  The observations and data analysis techniques are described in \S\ \ref{vla_obs}.  In \S\ \ref{vla_res}, the results from the total and polarized intensity images at 6 and 20 cm are presented, including a detailed discussion of the spectral index distribution of nonthermal radio filaments.  Section \ref{vla_dis} discusses the nature of the compact and extended sources in the survey, while conclusions are given in \S \ref{vla_con}.

\section{Observations and Data Reductions}
\label{vla_obs}
\subsection{VLA Data}
Observations with the VLA\footnote{The VLA is operated by the National Radio Astronomy Observatory, which is a facility of the National Science Foundation operated under cooperative agreement by Associated Universities, Inc.} at 6 and 20 cm were designed to have similar sensitivities, resolutions, and {\it uv} coverage, allowing unbiased spectral index measurements to be made.  Table \ref{vla_params} summarizes the parameters of the 6 and 20 cm surveys.  The 6 cm observations were conducted in the compact, hybrid, DnC configuration.  The 20 cm observations were conducted in three configurations:  the CnB configuration was observed for two days, the DnC configuration for one day, and D configurations for one day.  The ratios of the sizes of neighboring configurations (e.g., CnB and DnD) are similar to the ratios of the observing frequencies, such that observations have similar sensitivities to extended emission.  For the 6 and 20 cm observations, the default continuum observing set up was used, with two, 50-MHz bandpasses centered at (4.885, 4.835 GHz) and (1.465, 1.385 GHz), respectively.  Two of the 27 antennas were out of comission during all of the 6 and 20 cm observations.

\begin{figure}[tbp]
\begin{center}
\includegraphics[width=0.75\textwidth,bb=0 200 500 600]{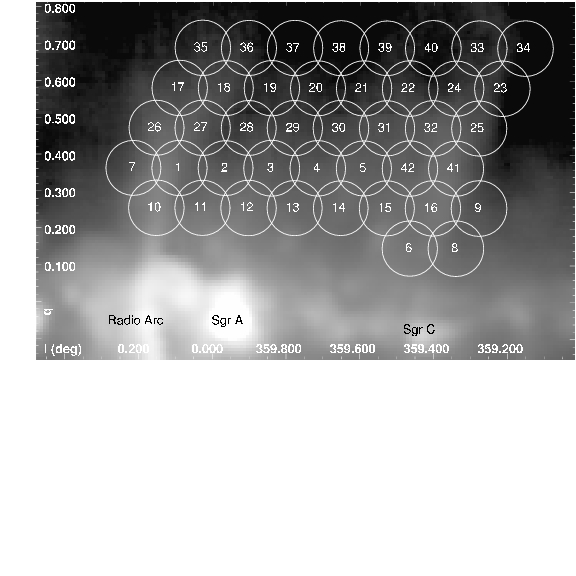}

\includegraphics[width=0.75\textwidth,bb=0 200 500 600]{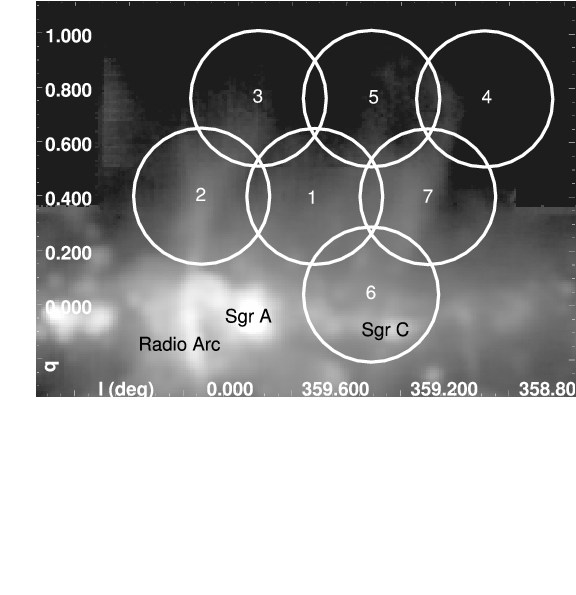}
\end{center}
\caption{\emph{Top}:  The coverage of the 6 cm, DnC-config VLA observations are shown in white circles.  The circles show the 42 pointings arranged in a hexagonal pattern with sizes representing the half-power of the primary beam of 9\arcmin.  A hexagonal pattern is used because it is efficient at covering an area with a circular beam and makes the sensitivity relatively uniform;  the sensitivity varied by about a factor of two across the inner part of this survey.  The gray scale shows a 6 cm, GBT continuum survey of the region, described in detail in \citet{gcsurvey_gbt}.  Field numbers show the order of observation. \emph{Bottom}:  Same as the top panel, but the circles show the locations of the seven, 20 cm, CnB-config pointings.  The sizes of the circles correspond to the half-power of the primary beam size of 30\arcmin. \label{coverage}}
\end{figure}

Figure \ref{coverage} shows the 6 and 20 cm survey regions, respectively.  In general, the surveys are designed to cover the Galactic center lobe (GCL), a possible signature of a mass-outflow that is discussed in more detail elsewhere \citep{s84,b03,thesis}.  The 6 cm fields also align at the southern edge with another 6 cm survey in the VLA archive.  The 20 cm data complement the high-resolution data of Paper I with a similar resolution and sensitivity, but better coverage at higher latitudes near the GC region.

Standard flux and phase calibration was done by observing 1331+305 (3C286) and 1751--253, respectively.  Phase calibration observations were done roughly every 24 minutes at 6 cm and every 40 minutes at 20 cm.  The phase calibrator was found to have a flux of 0.46 Jy and 1.13 Jy at 6 and 20 cm, respectively.  The polarization angle was calibrated by assuming a position angle of 66\sdeg\ for 1331+305.  To improve {\it uv} coverage, each pointing was observed in several intervals throughout the 7-hour observing period for a total integration time of roughly 30 and 110 minutes at 6 and 20cm, respectively.  The {\it uv} data were rotated into galactic coordinates prior to imaging.

\subsection{Imaging VLA Data}
At 6 cm, imaging was done using IMAGR in AIPS.  The compact source catalog was made from VLA data with {\it uv} spacings $>0.8$ k$\lambda$;  it was found that removing these short {\it uv} spacings simplified the detection of compact sources without significantly sacrificing overall sensitivity.  VLA images for studying extended structures and for combining with GBT images (see \S \ref{feathering} ) have no {\it uv} cut.  The data corresponding to Field 32 was self-calibrated using a $\sim$40-mJy source before the final images were made.  To study emission from the entire region surveyed, images were convolved to the minimum beam size shared by all 42 fields of 14\arcsec\ $\times$ 9\arcsec\ with $\theta_{PA}=70$\sdeg\ (East of celestial North; roughly 128\sdeg\ East of Galactic North) and mosaicked using FLATN.  Table \ref{image_resolutions} summarizes the resolutions of all images used in this work.

A similar process was used to clean stokes Q and U images for studying polarized emission at 6 cm.  Images used to study the polarized intensity and position angle were made from the average of the two 50 MHz bands, while images for studying the rotation measure used images from each band individually.  To test possible bias induced by the clean algorithm, the data were imaged with the multi-resolution and standard clean methods.  In general, the resulting images and their derived quantities are the same for both cleaning techniques, which suggests that the results are not very sensitive to the cleaning method.

At 20 cm, imaging was done using the VLALB task\footnote{VLALB is a part of the 4MASS software add-on to AIPS; \url{http://www.cv.nrao.edu/$\sim$bcotton}}.  VLALB iteratively images and self-calibrates the data using the imaging results for sources brighter than 40 mJy for further phase calibration.  As was done at 6 cm, two sets of images were created:  one set using all data to be used in the study of extended soruces and one optimized for compact sources, using {\it uv} spacings $>0.8$ k$\lambda$.  The imaging for the compact source detection used only the 20 cm data in CnB configuration;  this configuration has similar {\it uv} coverage as the 6 cm, DnC configuration data and is thus sensitive to similar spatial scales.  As done with the 6 cm maps, the seven fields were convolved to the beam size of the 6 cm survey (14\arcsec\ $\times$9\arcsec) with $\theta_{PA}=60$\sdeg and mosaicked prior compact source detection.  For most fields, Sgr A was imaged simultaneously with the rest of the field to reduce the affect of its sidelobes.  The stokes Q and U parameters were also imaged in the seven fields with the same {\it uv} range and weighting and made into a polarized intensity image for studying polarized sources.

\subsection{Feathering VLA and GBT data}
\label{feathering}
Maps were created at 6 and 20 cm using information from VLA and GBT \citep{gcsurvey_gbt} images using the ``feathering'' technique.  Feathering combines interferometric observations with single-dish observations in the {\it uv} plane which corrects the ``missing flux'' problem of interferometric images and improves the appearance of the image \citep{b79}.  The shortest baseline separation for the VLA C and D configurations is less than half the GBT diameter, which helps when determining the relative scaling of the two data sets \citep{s02}.  The technique combines the images in the {\it uv}-plane while using a mask to select the spatial frequencies that each image samples best.  In total, the images are Fourier-transformed, multiplied by the mask, combined, and then inverse-Fourier-transformed to create the final, feathered image.  The VLA images used for feathering were created with data weighted to reduce sidelobes;  this weighting reduces resolution, as shown in Table \ref{image_resolutions}.  The entire process is performed using a series of custom tasks written for the 4MASS add-on to AIPS (\url{http://www.cv.nrao.edu/$\sim$bcotton}).

\begin{figure}[tbp]
\begin{center}
\includegraphics[width=0.9\textwidth]{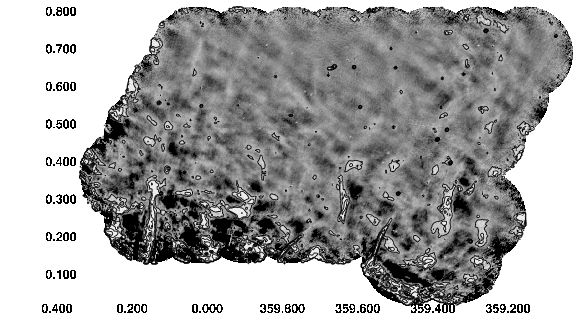}

\includegraphics[width=0.9\textwidth]{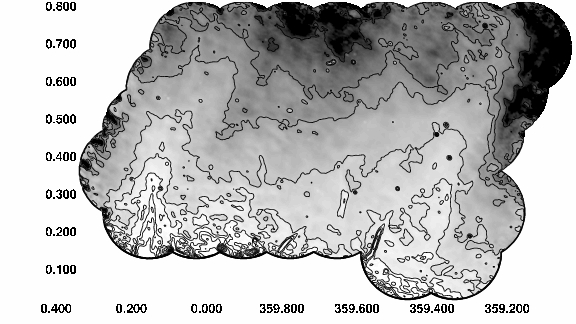}
\end{center}
\caption{\emph{Top}: Mosaic of 42, 6 cm VLA fields with Galactic coordinates.  Contours show flux density levels at $0.5 * 2^n$ mJy per 14\arcsec$\times$9\arcsec-beam, with $n=0-7$.  One $\sigma$ noise values in the mosaic image are typically about 0.1 mJy/beam, but range from 0.08 mJy beam$^{-1}$ (far from Sgr A*) to 0.5 mJy beam$^{-1}$ (near Sgr A*), as compared to a theoretical value of 0.03 mJy beam$^{-1}$.  The noise in the Q and U images were estimated by looking at the standard deviation outside of the primary beam.  The Q and U images have noise levels ranging from 50 to 120 $\mu$Jy (highest values near Sgr A*), which is a few times higher than the theoretical noise level of 30 $\mu$Jy.  \emph{Bottom}: Same data as above, but feathered with 6 cm GBT observations of the region and convolved to a beam size of 26\arcsec$\times$18\arcsec.  Contours show flux density levels of 1, 3, 5, 10, 15, 20, 25, 30, and 40 mJy beam$^{-1}$.  This image is used in the comparison of 6 and 20 cm fluxes of extended sources with slice analysis. \label{6large}}
\end{figure}

\begin{figure}[tbp]
\begin{center}
\includegraphics[width=0.75\textwidth]{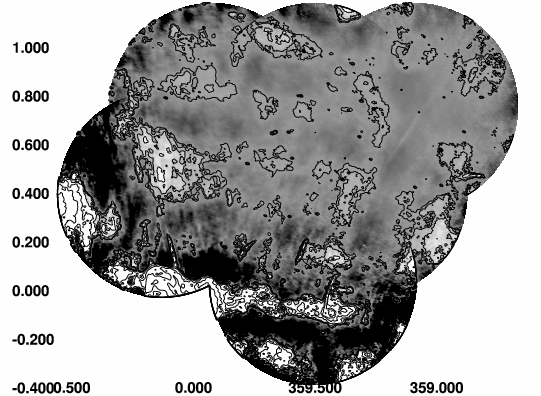}

\includegraphics[width=0.75\textwidth]{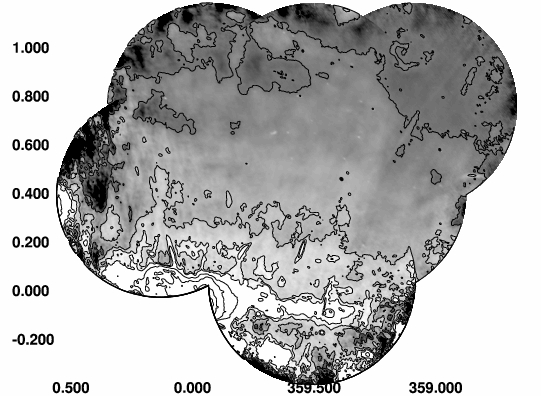}
\end{center}
\caption{\emph{Top}  Mosaic of the seven, 20 cm VLA fields with Galactic coordinates.  Contours show flux density levels at $2 * 2^n$ mJy per 14\arcsec$\times$9\arcsec-beam, with $n=0-7$.  Noise values for fields 1--7 are 0.8, 1.2, 0.6, 0.2, 0.5, 1.7, and 0.3 mJy/beam, respectively, as compared to a theoretical value of 0.02 mJy beam$^{-1}$.  \emph{Bottom}: Same data as above, but feathered with 20 cm, GBT observations of the region and convolved to a beam size of 26\arcsec$\times$18\arcsec.  Contours show flux density levels of 25, 50, 75, 100, 200, 400, 800, 1600, and 3200 mJy beam$^{-1}$.  This image is used in the comparison of 6 and 20 cm fluxes of extended sources with slice analysis. \label{20large}}
\end{figure}

Figures \ref{6large} and \ref{20large} show the distribution of 6 and 20 cm continuum emission on large scales with both VLA and VLA+GBT feathered maps.  The biggest difference between VLA and VLA+GBT feathered images is the strength of the continuum emission from the GCL, especially at 6 cm.  The feathered images also fill in the negative ``bowls'' in the interferometric images to produce a more uniform flux distribution near bright sources.  

\subsection{Polarized Continuum Mosaics}
\label{polnsec}
\begin{figure}[tbp]
\begin{center}
\includegraphics[width=0.9\textwidth]{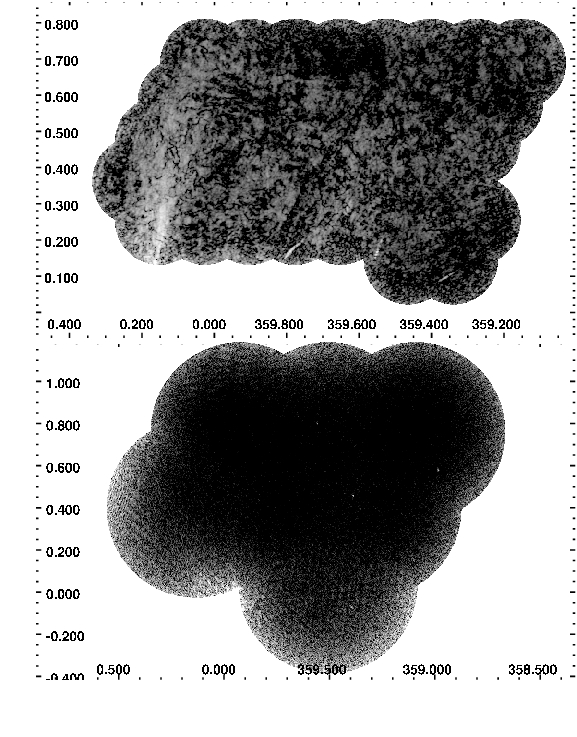}
\end{center}
\caption{\emph{Top}: Mosaic showing the 6 cm, polarized continuum toward the GC region.  The mosaic has a resolution of 15$\dasec5\times$13\arcsec.  Methods used to create this mosaic (and associated caveats) are described in \S \ref{polnsec}. \emph{Bottom}:  Same as the top panel, but showing the 20 cm, polarized continuum mosaic.  The mosaic has a typical resoultion of 11$\dasec5\times$8\arcsec. \label{poln}}
\end{figure}

A series of images were produced to study the polarized intensity and rotation measure structure toward filamentary and compact sources.  Figure \ref{poln} shows mosaics of the polarized continuum images at 6 and 20 cm.  As discussed below, the 6 cm maps seem to be filled with emission with typical brightness of $\sim0.5$ mJy and spatial structure on scales of 1--5\arcmin.  To improve the fidelity of the 6 cm images, those data were imaged with 30\% reduced sensitivity at 15 k$\lambda$ (using UVTAPER argument to IMAGR);  this produced images with a resolution of 15$\dasec5\times$13\arcsec.  At 20 cm, the maps were devoid of extended emission, so only high-resolution images were made, using the same techniques for the total intensity mosaics used for compact source detection;  this produced images with a resolution of 11\arcsec$\times$8\arcsec.

To create the polarized intensity mosaics, cleaned images were created of Stokes Q and U flux using IMAGR and mosaicked using FLATN.  The polarized intensity was created using the COMB task to correct for the noise bias, using the noise estimated from images outside the primary beam.  Mosaics of Stokes Q and U were also produced at each IF and combined to create polarization angle maps at two frequencies.  These images were used to estimate the rotation measure, as described below.

Wide-field, VLA imaging of polarized emission suffers from poorly understood systematic effects, so these mosaics should be interpreted with caution \citep{c94}.  An important systematic effect is the antenna-based polarization, which induces radially-oriented, linear polarization that increases with offset from the phase center.  There is no software or data available to properly correct for this effect, so the images presented here may suffer from this effect.  However, there are a few reasons that this instrumental effect may not be a concern in a few applications.  First, the instrumental polarization is strongest near the edge of each field, but mosaicking these fields will sometimes average biases with perpendicular orientations, thus cancelling the bias out.  Second, the parallactic angle for each source changes during the observation, which reduces antenna-based errors.  During the observations, the parallactic angle changed by about 80\sdeg, which reduces antenna-based polarization by about $1-\sin(\Delta\theta)/\Delta\theta\approx30$\%.  Finally we note that the analysis shown here focuses on the brightest polarized emission with counterparts in total intensity;  this polarized emission is much more than a few percent of the total intensity, which is the level at which these systematic effects are limited \citep{c94,c99}.  However, for much of the extended, 6 cm, polarized emission seen in Figure \ref{poln}, there is no total intensity counterpart, so it is more likely affected by instrumental effects;  discussion of this emission requires more rigorous analysis and thus is left for the future.

\section{Analysis and Results}
\label{vla_res}
\subsection{Extended Sources}
This section discusses the analysis and results for extended sources visible in the mosaicked images at 6 and 20 cm.  First, the spectral index and rotation measure analysis techniques are described, followed by results on individual objects.

\subsubsection{Spectral Index Analysis}
In order to investigate the distribution of spectral index across these sources, the spectral index was measured at several positions across each source.  The spectral index was measured from slices of the images, producing 1-d flux profiles across sources.  The spectral index is measured by comparing the background subtracted flux density of the sources in slices across the 6 and 20 cm feathered images.  The spectral index is measured by fitting a first-order polynomial to a source-free part of the slice to estimate the background and noise in the slice.  The images were convolved to a beam size of 26\arcsec$\times$18\arcsec before taking the slices and the spectral index is defined assuming $S_\nu\propto\nu^{\alpha}$.  The spectral index measurements were considered trustworthy only if they were found not to vary significantly with $\sim10$\% variations in the definition of the background region.  This work mostly presents spectral indices measured from slice analysis, since that technique makes it easier to measure the background and gives robust values than integrated analysis.  The technique is described in more detail elsewhere \citep{gcsurvey_gbt}.

In general, the spectral index measurements seem to be robust to observational effects.  To test the robustness of the slice technique to the slice orientation, pairs of perpendicular slices were made across a source and the resulting spectral index measurements were compared.  The index measured in these slices had equal spectral index measurements within their errors.  The indices measured on both feathered and VLA-only images were equal within their errors, except in 2 of the 74 slices compared.  There is no correlation between measurements of the spectral index and an image of the theoretical noise level for the 6 cm mosaic, which suggests that the noise and primary-beam correction do not bias the measurements.  There was also no correlation found between the spectral index uncertainty measured in the slices and the theoretical noise from the effect of the primary beam, suggesting that the noise is dominated by other effects.

Finally, we note that bandwidth smearing causes a radial smearing that reduces peak (per beam) brightness and can affect slice analysis.  Although bandwidth smearing is not commonly thought to affect extended sources, the widths of some NRFs are smaller than the beam in our observations, which means that smearing can affect their peak brightness in some cases.  In theory, bandwidth smearing increases with $\beta=\frac{\Delta\nu}{\nu_0}\frac{\theta_0}{\theta_{\rm{FWHM}}}$, where $\frac{\Delta\nu}{\nu}$ is the fractional bandwidth and $\frac{\theta_0}{\theta_{\rm{FWHM}}}$ is the offset from the phase center in beamwidths \citep{b99}.  This effect is only noticable in our 20 cm, where the primary beam size and bandwidth are relatively large.  The effects of bandwidth smearing depend on the location and orientation of the source, so they are described individually for each source.

\subsubsection{6 cm Rotation Measure Analysis}
The change in the polarization angle with wavelength due to Faraday rotation is a probe of the line of sight to the polarized source.  The amount of rotation across a given bandwidth is given by:
\begin{equation}
\Delta\theta = RM (\lambda_1^2-\lambda_2^2)
\end{equation}
\noindent where $RM$ is the integral of the product of the electron density and magnetic field along the line of sight.  Typically, observers use three or more wavelengths to measure the polarization position angle as a function of wavelength \citep[e.g.,][]{t86,h00,b05}.  These observations use the two, adjacent 50-MHz bands of the standard VLA observing configuration.  In this case, a position angle difference of one degree corresponds to a rotation measure of --220 rad m$^{-2}$, assuming no $2\pi$ phase wrap.  The calculation of an $RM$ from \dt assumes that the Faraday rotation has a $\lambda^2$ dependence, which is not always true \citep{b66, b05}.  Thus, in this work we assume a $\lambda^2$ dependence and give \emph{effective} $RM$ values, or \rmeff.

Studying Faraday rotation across two, adjacent 50-MHz IFs is not common \citep[for one example see][]{y88}, but there are a few reasons why it can give useful results in this application.  First, the $RM$ expected in the GC region covered by this survey is typically less than 2000 rad m$^{-2}$, with a maximum of about 3000 rad m$^{-2}$ in the Galactic plane \citep{y84,t86,t95,r05}.  For $RM=2000$ rad m$^{-2}$, the change in polarization angle between the two IFs of the present survey is about 9\sdeg.  The degeneracy in position angle for $\Delta\theta>180$\sdeg is unlikely in the present data, since this would require $RM$ values twenty times larger than is average ($RM>4\times10^4$ rad m$^{-2}$).  Secondly, the bandwidth depolarization expected in this data is roughly $1-\sin(\Delta\theta)/\Delta\theta=0.4$\% for $RM$ of 2000 rad m$^{-2}$, which shouldn't significanly affect the results.  However, it is worth noting that some areas of the polarized maps are completely depolarized;  this highlights the fact that these observations are not sensitive to $RM$ greater than $4\times10^4$ rad m$^{-2}$.  The opposite can also happen, since the total intensity is more likely to be resolved out than the polarized intensity, which tends to be broken into many small pieces by depolarization canals;  in these cases, the polarized fraction can exceed 100\%.

\subsubsection{Results for Nonthermal Radio Filaments}
\label{nrfsec}
\begin{figure}[tbp]
\includegraphics[width=\textwidth]{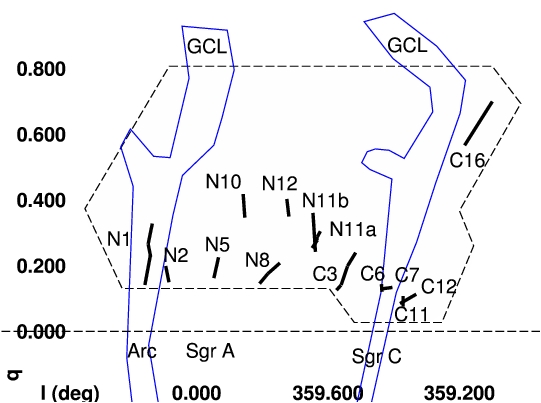}
\caption{Schematic of the NRFs and other major features of the GC region.  Each NRF detected at 6 cm is labeled in the figure.  The outline of the radio continuum emission from the GCL is shown in blue.  The extent of the 6 cm survey region is shown with the enclosed dashed region, while the Galactic plane is shown with a dashed line. \label{vla_nrfschem}}
\end{figure}

Several filamentary structures are apparent in the 6 and 20 cm maps.  Figure \ref{vla_nrfschem} shows a schematic view of the survey region with major features and filaments labeled.  Table \ref{nrftab} summarizes the properties of the known and candidate NRFs from the 6 cm data, including the 6/20 cm spectral index and polarized emission from several of the filaments, which confirms their identification as NRFs.  

\begin{figure}[tbp]
\includegraphics[width=6.5in]{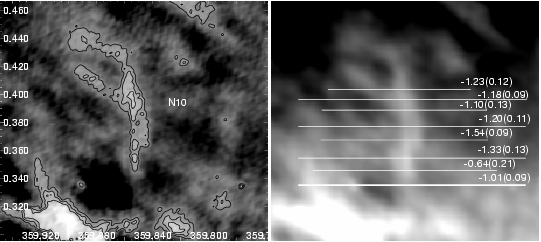}
\includegraphics[width=6.5in]{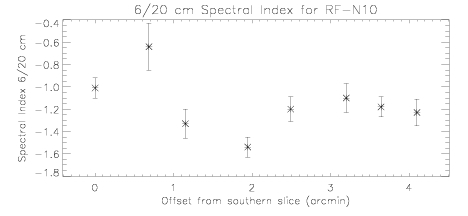}
\caption{\emph{Top left}:  High-resolution, VLA-only image of the N10 radio filament at 6 cm.  The image has a resolution of 15\arcsec$\times$8\arcsec\ and contour levels at 0.4, 0.6, and 0.8 mJy beam$^{-1}$.  \emph{Top right}:  Feathered, 6 cm image of the region, convolved to the resolution of the 20 cm feathered image of 26\arcsec$\times$18\arcsec.  The lines show the positions of slices across the 6 and 20 cm images for the spectral index study.  The measured value and error of $\alpha_{6/20}$ is shown near each slice.  \emph{Bottom}:  Spectral index measured for each slice shown in the right image.  The thick slice in the right image shows the southernmost slice, which measures the spectral index at the origin of the plot at (l,b) = (359\ddeg851, 0\ddeg336);  other measurements are along a line oriented 2\sdeg East of Galactic North. \label{n10fig}}
\end{figure}

Below, the analysis of the morphology, spectral index distribution and polarization properties of several filaments is discussed.  The format of Figure \ref{n10fig} is used throughout this work, as it connects the high-resolution and feathered images in the top panels to the accompanying spectral index analysis in the bottom panel.  Names of the filaments are given in Galactic coordinates with the designation of Paper I in parentheses.  The C1, N5, and N12 filaments are not discussed because they are only marginally detected, or are too near the survey edge to measure reliable properties.

$G359.85+0.39  (RF-N10)$ ---  This filament is described first, because it can be compared with previous work as a validation of the analysis method.  This filament was well studied by \citet{l01}, who observed it at 6, 20, and 90 cm and discovered its unusual, braided subfilaments.  Figure \ref{n10fig} shows a 6 cm, high-resolution image and the spectral index distribution across the N10 radio filament, which is located relatively far north of the Galactic plane and oriented nearly perpendicular to it.  We confirm that N10 is composed of at least two, parallel, unresolved components at 6 cm.  This filament has an integrated 6 cm flux density of 42.9$\pm$9.0 mJy and is polarized at 6 cm.

The spectral index changes observed in the feathered images of N10 are shown in Figure \ref{n10fig}.  The index changes significantly across the filament, ranging from --0.6 to --1.5, but with no regular trend.  The range of values and the lack of a trend measured here are consistent with the 6/20 cm spectral index behavior reported by \citet{l01}.  That work reported on a spatial gradient in the 20/90 cm index, but noted that the 6/20 cm index shows no similar changes.  Those authors attributed the difference to missing flux at short spacings, although the present observations should be sensitive to emission on all size scales.  The consistency between the present work and \citet{l01} give some confidence in the spectral index analysis.

\begin{figure}[tbp]
\includegraphics[scale=0.44]{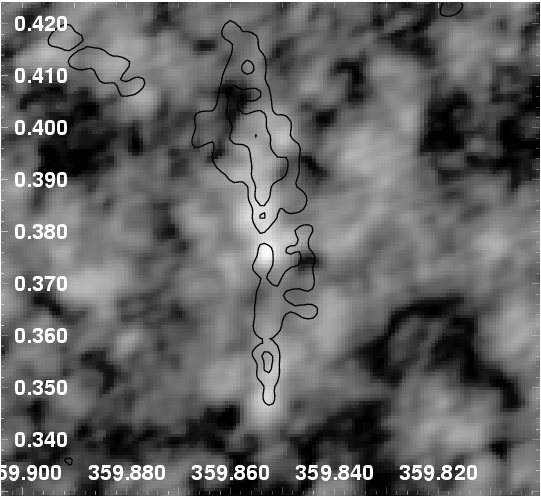}
\hfil
\includegraphics[scale=0.44]{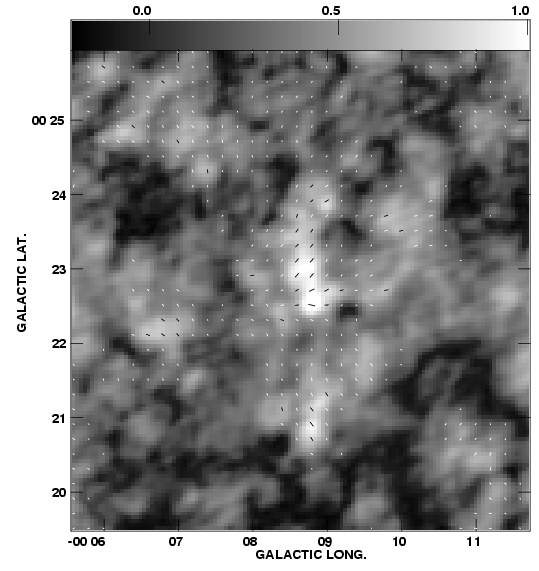}
\caption{\emph{Left}:  Gray scale shows the 6cm polarized intensity from RF-N10 (G359.85+0.39) for brightnesses ranging from -1 to 5 mJy beam$^{-1}$.  The contours show 6 cm total intensity at levels of 0.5 and 0.7 mJy beam$^{-1}$.  \emph{Right}:  Image of the same region with the vectors showing the 6 cm polarization angle averaged over the 100 MHz band.  Vectors are shown for regions with polarized intensity brighter than 0.2 mJy beam$^{-1}$. \label{n10pol}}
\end{figure}

The polarized emission from nonthermal filament G359.85+0.39 is shown in Figure \ref{n10pol}.  The polarized brightness is irregular, but does show some correlation with total intensity emission.  The polarized intensity is found in a few isolated islands of emission along the brightest subfilament.  The polarized structure and its correlation with the total intensity is similar to that observed by \citet{l01}.  The brightest polarized emission is about 1.2 mJy beam$^{-1}$ ($\sim10\sigma$ significance) and the polarization fraction is larger than 100\% at a few locations along the brightest subfilament.  The average \dt for the region with polarized emission is $-1\pm1$\sdeg, or \rmeff$=220\pm220$ rad m$^{-2}$.

\begin{figure}[tbp]
\includegraphics[width=6.5in]{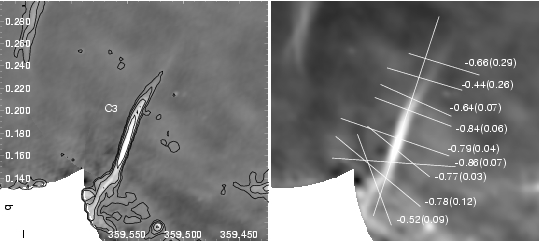}
\includegraphics[width=6.5in]{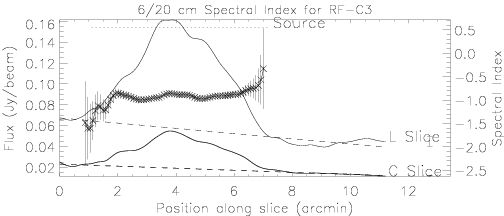}
\caption{Same as Figure \ref{n10fig}, but for the C3 radio filament.  Contours levels are at 1, 2, 4, and 8 mJy beam$^{-1}$ and the top left image has a beamsize of 15\arcsec$\times$10\arcsec.  The bottom plot shows the 6 and 20 cm flux densities (labeled by their band names, ``C'' and ``L'', respectively) with their spectral index as measured by a single slice along the long axis of the filament.  The plot shows the spectral index at each point along the filament, starting from the southern end of the slice.  The slice begins at (l,b) = (359\ddeg571, 0\ddeg106) and is oriented at 342\sdeg East of Galactic North.  The best-fit, first-order polynomial background is shown with a dashed line for both flux slices.  The background is fit ignoring the portion of the slice labeled ``source''.  \label{c3fig}}
\end{figure}

$G359.54+0.18 (RF-C3)$ --- The 6 cm VLA image and the spectral index distribution of the C3 filament, also known as ``the ripple'', are shown in Figure \ref{c3fig}.  This filament was found by \citet{b89}, who noted how it bends where it apparently is interacting with a molecular cloud \citep{b89,r03}.  This filament is one of the longest ($\sim11$\arcmin) and brightest at 6 cm (integrated flux density $490\pm60$ mJy, consistent with previous observations \citep{b89}.

The spectral index of the filament was measured with both perpendicular slices (as for RF-N10) and with a single slice along its long axis;  the index from the single slice along its long axis is shown in the bottom panel of Figure \ref{c3fig}.  The spectral index is relatively constant across C3, with a typical value of $\alpha_{LC}=-0.8\pm0.05$.  There is some suggestion in the slice analysis that the spectral index is flatter in the fainter portions near the northern and southern ends of C3.  The integrated spectral index was measured previously between 90 and 20 cm as $-0.8\pm0.1$ \citep{a91} and 20 and 6 cm as $\sim-0.8$ \citep{y05}, consistent with the present slice analysis.  However, the intergrated spectral index between 6 and 3.6 cm steepens significantly to $\sim-2$ \citep{y05}.

\begin{figure}[tbp]
\includegraphics[scale=0.44]{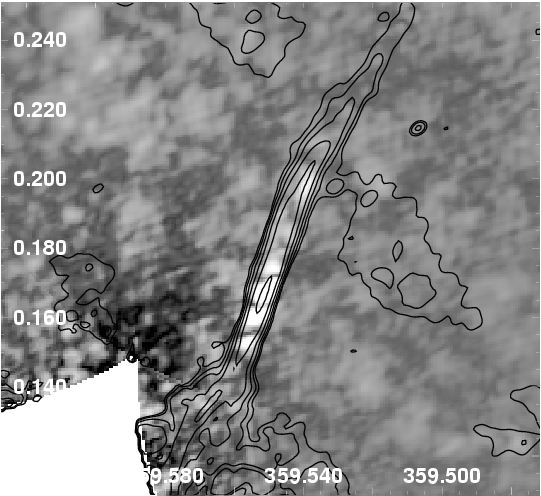}
\includegraphics[scale=0.44,angle=270,bb=500 0 600 500]{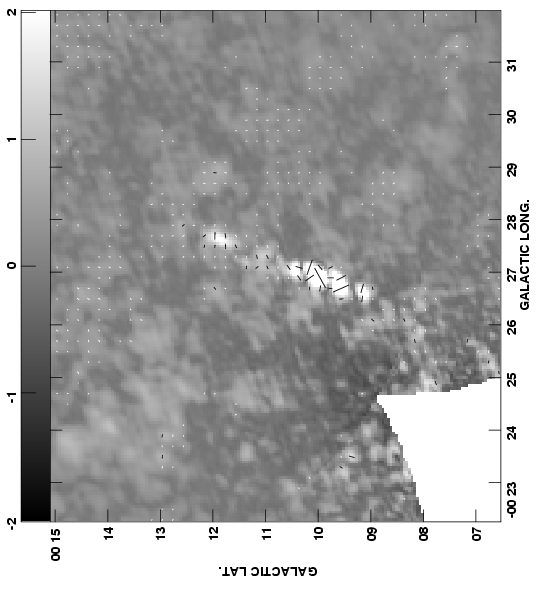}
\caption{\emph{Left}: Gray scale shows the 6cm polarized intensity from RF-C3 (G359.54+0.18) for brightnesses ranging from -1 to 5 mJy beam$^{-1}$.  The contours show 6 cm total intensity at levels of $0.5*2^n$ mJy beam$^{-1}$, for $n=0-5$. \emph{Right}: Image of the same region with the vectors showing the 6 cm polarization angle averaged over the 100 MHz band.  Vectors are shown for regions with polarized intensity brighter than 0.2 mJy beam$^{-1}$. \label{c3pol}}
\end{figure}

Figure \ref{c3pol} shows a clear 6 cm, polarized counterpart to RF-C3, as had been noted before \citep{b89,y97}.  The polarized emission is broken into roughly five ``islands'' of polarization about 30\arcsec\ across, separated from each other by depolarized canals.  Four of the islands are located along the brightest total intensity emission where the NRF is straight, while the fifth island is north of the others, near the northern bend in the NRF.  The brightest polarized intensity from RF-C3 is 6 mJy beam$^{-1}$ with a polarization fraction of about 35\%.  The other polarized islands have lower polarized intensities with polarization fractions ranging from 20 to 30\%.

Each of the polarized islands has a relatively constant \dt, so an error-weighted average of \rmeff\ is calculated for each one.  While averaging is not strictly appropriate for nonlinear quantities like angles, it is approximately correct in the limit of small angles (\dt $\ll 1$ rad $\approx 57$\sdeg).  From south to north, the average \dt value for the islands are:  11$\pm$6\sdeg$=-2420\pm1320$ rad m$^{-2}$ (at 359\ddeg556,0\ddeg152), $7\pm3$\sdeg$=-1540\pm660$ rad m$^{-2}$ (at 359\ddeg554,0\ddeg163), $10\pm2$\sdeg$=-2200\pm440$ rad m$^{-2}$ (at 359\ddeg550,0\ddeg167), $18\pm5$\sdeg$=-3960\pm1100$ rad m$^{-2}$ (at 359\ddeg549,0\ddeg174), and $15\pm4$\sdeg$=-3300\pm880$ rad m$^{-2}$ (at 359\ddeg539,0\ddeg198).

The morphology seen in the present survey is similar to that of \citet{y97}, which presented detailed study of the polarization properties of RF-C3 at 6 and 3.6 cm.  The clump at RA,Dec (B1950, as used in \citet{y97}) = (17:40:41,-29:12:30) has $RM\approx-2700$ rad m$^{-2}$, compared to $\Delta\theta\approx18\pm5$\sdeg\ $\approx-3960\pm1100$ rad m$^{-2}$ in the present survey.  Another clump, at (17:40:43,-29:12:40), has $RM\approx-2000$ rad m$^{-2}$, compared to $\Delta\theta\approx10\pm2$\sdeg\ $\approx-2200\pm440$ rad m$^{-2}$ in the present survey.  A third clump, at (17:40:44,-29:12:45), has $RM\approx-1500$ rad m$^{-2}$, compared to $\Delta\theta\approx7\pm3$\sdeg\ $\approx-1540\pm660$ rad m$^{-2}$ in the present survey.  Thus, in general, there is good agreement between the \rmeff\ of the present survey and the more rigorously measured values of \citet{y97}.

\begin{figure}[tbp]
\includegraphics[width=6.5in]{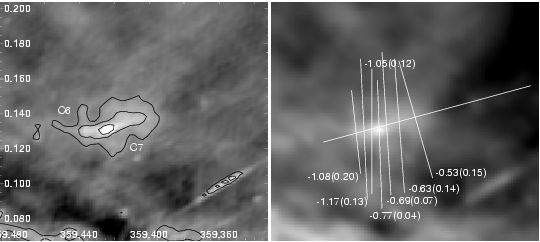}
\includegraphics[width=6.5in]{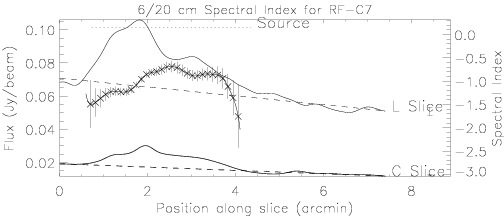}
\caption{Same as Figure \ref{n10fig}, but for the C6 and C7 radio filaments. Contour levels are at 1, 2, and 4 mJy per 15\arcsec$\times$10\arcsec-beam.  The C11/C12 filament complex is also seen in the southwest of the image.  The plot shows the 6 and 20 cm flux densities (labeled by their band names, ``C'' and ``L'', respectively) with their spectral index as a function of position along the length of the filament;  the slice starts at (l,b) = (359\ddeg455, 0\ddeg124) and is oriented 285\sdeg East of Galactic North. \label{c7fig}}
\end{figure}

$G359.42+0.13 (RF-C7)$ --- Figure \ref{c7fig} shows the 6 cm continuum and spectral index distribution of the C7 radio filament.  The filament is relatively straight and short (3\arcmin), oriented parallel to the Galactic plane with a significant brightening at its midpoint.  The integrated 6 cm flux density is around 28.2$\pm$8.3 mJy and no polarized emission is found toward this filament brighter than a confusion limit of 0.4 mJy beam$^{-1}$.  Using slices perpendicular to the filament and along its length, we find a trend for steepening toward its eastern half, with values changing from $\sim-0.7$ to $\sim-1.1$.  The spectral index changes near the brightest portion of the filament, so errors in background subtraction should be minimal;  also, the consistency of the two different slice directions adds confidence to this spectral index change.  The spectral index change can also be seen by visually comparing the change in flux about the filament midpoint at 20 cm (changes by roughly a factor of two) and at 6 cm (roughly unchanged).  

$G359.44+0.14 (RF-C6)$ --- This source is detected in 6 cm images (see Fig. \ref{c7fig}) at the eastern edge of the C7 filament, but is not seen in the 20 cm images presented here.  The filament is very faint and short, with a peak 6 cm flux density of 1.5$\pm$0.1 mJy beam$^{-1}$ and a length of about 2\arcmin.  The relative alignment of the C6 and C7 filaments is similar to that of nearby filaments C11 and C12, as was noted in Paper I.  These two groups of filaments have pairs of filaments with one oriented parallel and one perpendicular to the Galactic plane;  in both cases, the filament parallel to the plane is the brighter and longer of the pair.

\begin{figure}[tbp]
\includegraphics[width=6.5in]{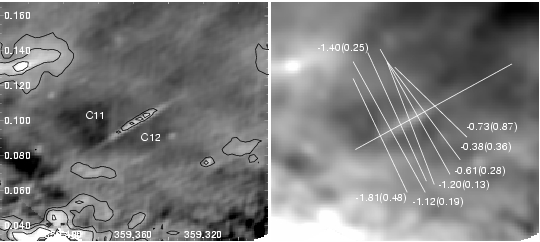}
\includegraphics[width=6.5in]{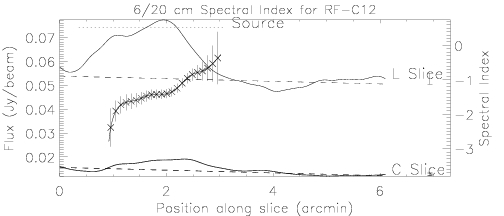}
\caption{Same as Figure \ref{n10fig}, but for the C11 and C12 radio filaments. Contour levels are at 0.5, 1, and 2 mJy per 15\arcsec$\times$10\arcsec-beam.  The plot shows the 6 and 20 cm flux densities (labeled by their band names, ``C'' and ``L'', respectively) with their spectral index as a function of position from the southern end of the filament, starting at (l,b) = (359\ddeg388, 0\ddeg083), along a line oriented 298\sdeg East of Galactic North. \label{c12fig}}
\end{figure}

$G359.36+0.10 (RF-C12)$ --- Figure \ref{c12fig} shows an overview of the radio continuum emission of the C12 filament, which was discovered at 90 cm by \citet{n04}.  It has a length of about 4\arcmin, a relatively constant brightness, and no significant curvature.  The filament is oriented at an angle of about 30\sdeg\ relative to the plane.  

These observations are the first 6 and 20 cm observations done for this C12.  The spectral index values for C12 were measured with perpendicular and parallel slices, as was done for C7.  Both sets of slices show a significant steepening in $\alpha_{LC}$ toward its eastern end.  This is also visible in Figure \ref{c12fig}, which shows that there is significant 20 cm emission near the eastern edge of the filament, but no 6 cm emission and that the peaks of the 6 cm and 20 cm emission are shifted relative to each other;  the physical offset of the 6 and 20 cm-emitting regions is similar to that observed by \citet{l01} for RF-N10.  The morphology, brightness, and spectral index changes across this filament are similar to that seen in the C7 filament.

\begin{figure}[tbp]
\includegraphics[scale=0.44]{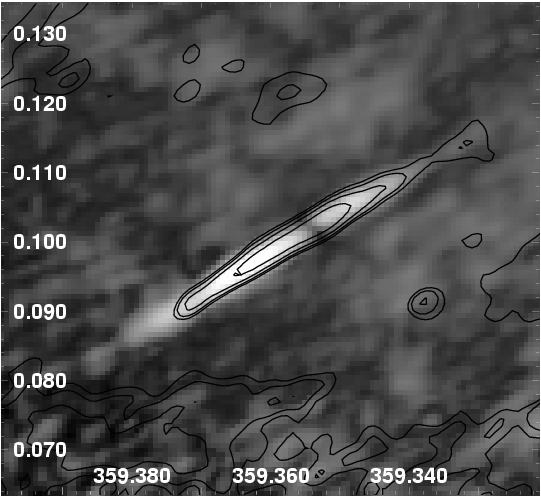}
\hfil
\includegraphics[scale=0.44,angle=270,bb=500 0 600 500]{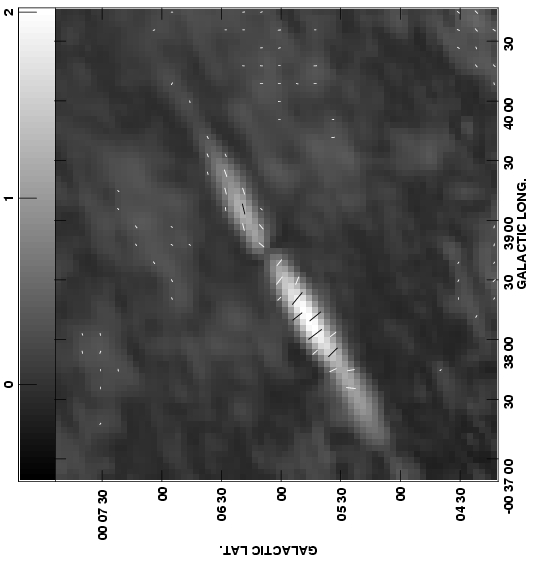}
\caption{\emph{Left}: Gray scale shows the 6cm polarized intensity from RF-C12 (G359.36+0.10) for brightnesses ranging from -1 to 5 mJy beam$^{-1}$.  The contours show 6 cm total intensity at levels of $0.25*2^n$ mJy beam$^{-1}$, for $n=0-3$. \emph{Right}: Image of the same region with the vectors showing the 6 cm polarization angle averaged over the 100 MHz band.  Vectors are shown for regions with polarized intensity brighter than 0.2 mJy beam$^{-1}$. \label{c12pol}}
\end{figure}

As seen in Figure \ref{c12pol}, polarized emission is seen across the entire 4\arcmin\ length visible in 6 cm total intensity.  This is the first detection of linearly-polarized emission from G359.36+0.10 and confirms that it is a NRF.  There is one depolarization canal near the midpoint of the filament that divides it into eastern and western halves.  The peak, 6 cm polarized intensity of about 2 mJy beam$^{-1}$ is located along the eastern half, while the peak in 6 cm total intensity is located along the western half.  Thus, the polarization fraction changes along the length of RF-C12 with a value of 30\% at the point of brightest total intensity and 70\% at the point of hightest polarized intensity.  The polarized emission is clearly detected at the faintest ends of the filament in total intensity such that the apparent polarization fraction in those regions is higher than 100\%.  

The \dt observed in the polarized emission of RF-C12 gives an estimate of the $RM$ toward the filament.  The eastern and western halves have averages of \dt$\approx-4\pm2$\sdeg\ and $-2\pm2$\sdeg, respectively, giving \rmeff$=880\pm440$ rad m$^{-2}$ and $440\pm440$ rad m$^{-2}$.

$G359.37+0.11 (RF-C11)$ --- This source is detected in 6 cm images at the eastern edge of C12 (see Fig. \ref{c12fig}), but is not seen in the 20 cm images presented here.  It is one of the faintest (1$\pm$0.2 mJy beam$^{-1}$) and shortest (2\arcmin) filaments observed in this survey.  As noted above, this filament and its relation to the C12 filament is similar to the C6/C7 filament complex in their orientation and relative brightness.

\begin{figure}[tbp]
\begin{center}
\includegraphics[width=6in]{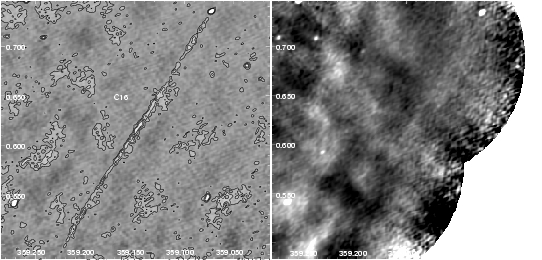}
\end{center}
\caption{The left image shows a high-resolution, 20 cm image around the C16 radio filament.  The 20 cm contour levels are at 0.3, 0.6, and 0.9 mJy per 14\arcsec$\times$9\arcsec-beam. \label{c16fig}}
\end{figure}

$G359.21+0.54 (RF-C16)$ --- Figure \ref{c16fig} shows the 20 cm continuum and spectral index distribution for the C16 filament, first detected at 90 cm by \citet{n04}.  The present survey provides the first observations at 6 cm and the highest latitude coverage at 20 cm.  This filament is unusual in that it is very long ($\sim$10\arcmin) and extends far from the Galactic plane ($0\ddeg6\approx80$ pc, in projection).  If confirmed as an NRF, this would increase the latitude range occupied by NRFs dramatically, out to (l,b) = (359\ddeg1,0\ddeg75).  This filament is unlikely to be an artifact, since it appears in two separate fields of the 20 cm survey.  The C16 filament is also interesting because it is oriented nearly parallel to the GCL in this region, as seen on large scales in Figure \ref{20large}.  There is some indication of a second filament, or an extended region of emission, parallel and just west of the brightest part of C16.  No polarized emission is evident at 6 or 20 cm to a level of 0.4 and 0.3 mJy beam$^{-1}$, respectively.

The filament is only marginally detected at 6 cm, so no spectral index analysis is done.  Nominally the 6 cm peak brightness is detected with roughly 2$\sigma$ confidence, but this low-level emission appears to be extended along the 20 cm feature for a length of 3\arcmin;  a rough upper limit on the 6/20 cm spectral index is $\alpha_{LC}<-1.2$.

There is a compact source detected at 6 and 20 cm that happens to lie along the C16 filament, with RA,Dec (J2000) = (17:40:31.410,--29:20:19.11) and visible at the top right of Figure \ref{c16fig}.  Comparing the density of compact sources in the region to the apparent area covered by the filament, implies a roughly 1 in 4 chance that this alignment is coincidental.

\begin{figure}[tbp]
\includegraphics[width=6.5in]{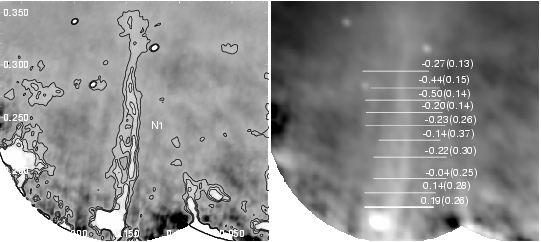}
\includegraphics[width=6.5in]{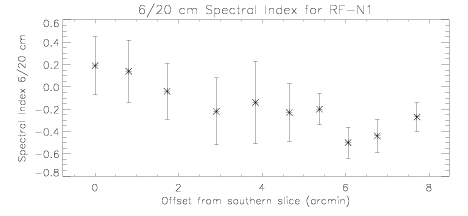}
\caption{Same as Figure \ref{n10fig}, but for the N1 radio filament.  The \hii\ region G0.17+0.15 is visible at the bottom of the images.  The contour levels are at 1, 2, and 3 mJy per 15\arcsec$\times$10\arcsec-beam.  The thick slice in the right image shows the southernmost slice, which is also the origin of the plot, at (l,b) = (0\ddeg152, 0\ddeg166);  the spectral index is measured along a line oriented 357\sdeg East of Galactic North. \label{n1fig}}
\end{figure}

$G0.15+0.23 (RF-N1)$ --- Figure \ref{n1fig} shows the N1 filament.  This filament is a part of emission from the Radio Arc, the brightest and most complex collection of NRFs known \citep{y84,m96}.  The N1 filament is north of the brightest part of the Radio Arc, but is contiguously connected to that region (to the south) and the GCL (to the north) in images of high-resolution polarized emission \citep{y88}.  This filament seems to bend around the \hii\ region at the bottom of the figure (known as G0.17+0.15), which suggests that these nonthermal structures interact with star-forming regions or their associated molecular and ionized gas \citep[][Paper I]{s94}.  The 6 cm data show for the first time that there is significant polarized emission associated directly with the N1 filament, confirming that it is an NRF (polarized emission fills this region, but has never previously been definitively associated with N1).

The N1 filament spectral index between 6 and 20 cm is studied by taking slices across the images, as described before.  Note that N1 is embedded in diffuse nonthermal emission, which the slice analysis subtracts in order to calculate $\alpha_{LC}$.  Figure \ref{n1fig} shows the spectral index steepening steadily toward the northernmost slices, with values ranging from +0.2 to --0.5.  The relatively flat spectral index is unusual for NRFs in general, but not so for the brightest NRFs, like in the Radio Arc, which has a spectral index $\sim0$ for GHz frequencies \citep{y86,p92}.  Interestingly, the spectral index does show moderate steepening toward the north;  this trend is seen in the large-scale emission of the GCL, as observed by the GBT \citep{p92,thesis}.

\begin{figure}[tbp]
\includegraphics[scale=0.44]{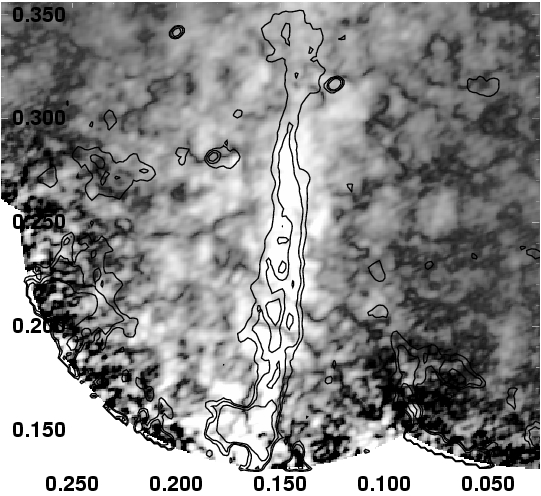}
\hfil
\includegraphics[scale=0.44,angle=270,bb=500 0 600 500]{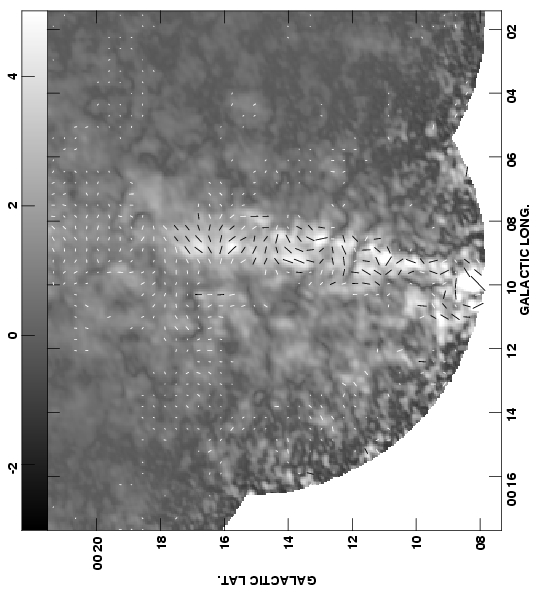}
\caption{\emph{Left}:  Gray scale shows the 6cm polarized intensity from RF-N1 (G0.15+0.23) for brightnesses ranging from -1 to 5 mJy beam$^{-1}$.  The contours show 6 cm total intensity at levels of 1, 2, and 4 mJy beam$^{-1}$.  \emph{Right}:  Image of the same region with the vectors showing the 6 cm polarization angle averaged over the 100 MHz band.  Vectors are shown for regions with polarized intensity brighter than 1 mJy beam$^{-1}$. \label{n1pol}}
\end{figure}

The polarized emission from RF-N1 is shown in Figure \ref{n1pol}.  The region is filled with clumpy polarized emission formed by depolarization canals.  The brightest polarized intensity is about 5.5 mJy beam$^{-1}$.  In many locations, the polarized brightness is greater than the total intensity, so the polarized fraction exceeds 100\%.  Although RF-N1 is in the midst of this extended polarized emission, this image is the first time polarized emission was specifically associated with the filament.  The average \dt for RF-N1 was measured for $b=0\ddeg16-0\ddeg30$ to be $-2\pm0\ddeg5$, equivalent to \rmeff$=440\pm110$ rad m$^{-2}$.

\begin{figure}[tbp]
\includegraphics[width=6.5in]{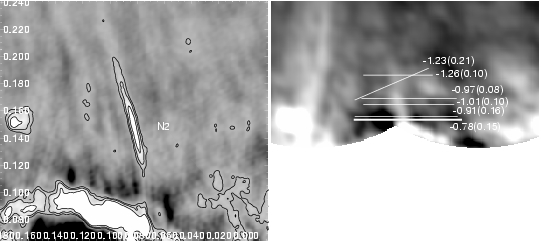}
\includegraphics[width=6.5in]{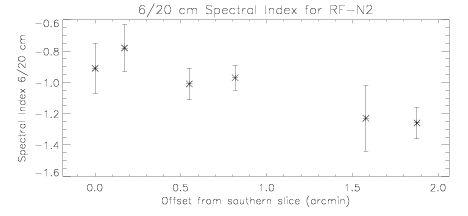}
\caption{Same as Figure \ref{n10fig}, but for the N2 radio filament.  The left image shows 20 cm brightness with contours levels at 5, 10, and 20 mJy per 14\arcsec$\times$9\arcsec-beam.  The thick slice in the right image shows the southernmost slice, which is also the origin of the plot, at (l,b) = 0\ddeg084, 0\ddeg152);  the spectral index is measured along a line oriented 14\sdeg East of Galactic North.  \label{n2fig}}
\end{figure}

$G0.08+0.15 (RF-N2)$ --- Figure \ref{n2fig} shows the 20 cm emission and spectral index distribution along the N2 radio filament (also called the ``Northern Thread''.  \citet{l99} extensively studied this long (15\arcmin), vertical filament and discovered its linearly polarized emission at 3.6 and 6 cm.  This filament is located at the edge of the 6 cm survey region, so only a fraction of the filament is seen and no integrated 6 cm flux densities are given in Table \ref{nrftab}.  

The spectral index measured in slices of N2 show a clearly nonthermal spectral index that seems to steepen at higher latitudes.  However, since the primary beam corrections for the 6 cm VLA data are large at the edge, it creates strong variation in the background of the feathered data used here.  Also, primary-beam corrections at 6 cm vary by a factor of 2--3 across N2, so it is possible that the spectral index at the edge is affected by uncertainties in this correction.  The spectral index of N2 has been measured south of the region covered by the present survey to have a flatter spectral index with little spatial variation, ranging from 0.0 to --0.5 \citep{l99}\footnote{Note the spectral index measured in Figure 11 of \citet{l99} is $\sim230$\arcsec\ south of our measurements.}.  Previous 20/90 cm spectral index measurements found typical values around $-0.6\pm0.1$, with no significant change across a 4\arcmin\ length \citep{a91,l00}.  However, those measurements are dominated by emission south of the present ones, since the N2 filament is brightest near the middle of its 10\arcmin length and the present work only detects the northern 2--3\arcmin.

There are two islands of enhanced 6cm, polarized emission coincident with the total intensity emission from N2.  This is a marginal detection and the images are noisy, since the filament is at the edge of the survey area.  However, polarized emission was previously detected and described in detail by \citet{l99}.


$G359.96+0.09 (RF-N5)$ --- This filament, also known as RF-N5 (Paper I) and ``the Southern Thread'' \citep{l99}, has a length of roughly 3\arcmin\ and is oriented perpendicular to the Galactic plane.  No figure is shown for this filament, since it is at the edge of our survey region and has a 6 cm surface brightness of RF-N5 is less than 1 mJy beam$^{-1}$.  There is some weak polarized emission associated with the filament, but it is not very significant and is well studied in \citet{l99}.

\begin{figure}[tbp]
\includegraphics[width=6.5in]{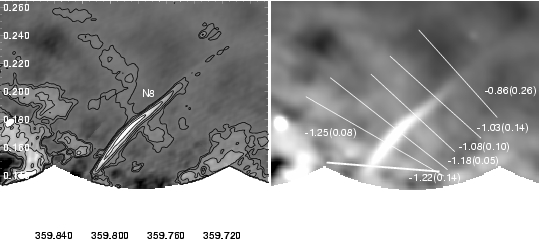}
\includegraphics[width=6.5in]{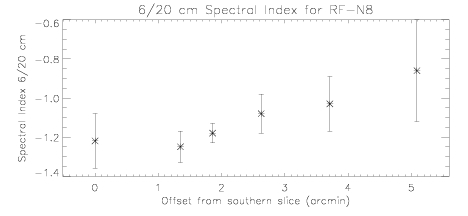}
\caption{Same as Figure \ref{n10fig}, but for the N8 radio filament.  The left image has contour levels at 0.5, 1, 2, 4, and 8 mJy per 15\arcsec$\times$10\arcsec-beam.  The thick slice in the right image shows the southernmost slice, which is also the origin of the plot, at (l,b) = (359\ddeg808,0\ddeg148);  the spectral index is measured along a line oriented 315\sdeg East of Galactic North.  \label{n8fig}}
\end{figure}

$G359.79+0.17 (RF-N8)$ --- First identified in wide-field, 20 and 90 cm observations \citep{y86,a91}, N8 was typical of the brightest filaments known in that it was several arcminutes long and had a slightly wavy shape.  Figure \ref{n8fig} shows that the first 6 cm image of N8.  North of the brightest emission the curve opens toward and then away from the plane again in the faintest northern parts of the filament.  As in previous observations, subfilamentation is apparent in the faint northern and southern portions of N8.  The 20 cm image shows a broader, fan-like shape in the southern part of the filament (see also Fig. 19b of Paper I).  This filament has an integrated 6 cm flux density of $226.7\pm24.8$ mJy and was previously found to have polarized radio continuum emission, which is also found in the present 6 cm observations \citep{l99,la00}.

These 6 and 20 cm observations provide us with the first chance to study the spectral index and its changes at these wavelengths.  The spectral index found by slice analysis shows a slight flattening in $\alpha_{LC}$ toward the west, ranging from --1.3 to --0.9.  Previous work found an integrated spectral index between 90 and 20 cm of $-0.6\pm0.1$ \citep{a91}.  The N8 index flattens at higher latitudes, which is the opposite latitude dependence as seen in N1 and N2, but similar to that seen in C3 and C12.  The N8 filament is also similar to C3 and C12 (and different from N1 and N2) in the angle that the filament makes with the Galactic plane.

However, bandwidth smearing may bias the spectral index measured for the N8 and N10 filaments.  The bandwidth-smearing bias for slice analysis was estimated by comparing the slice and integrated flux densities for the nearby compact source, G359.872,+0.178;  this finds a bias of $\Delta\alpha\sim+0.4$ for the compact source.  This bias is an upper limit to that in N8 and N10, since the filaments are not unresolved.  

\begin{figure}[tbp]
\includegraphics[scale=0.44]{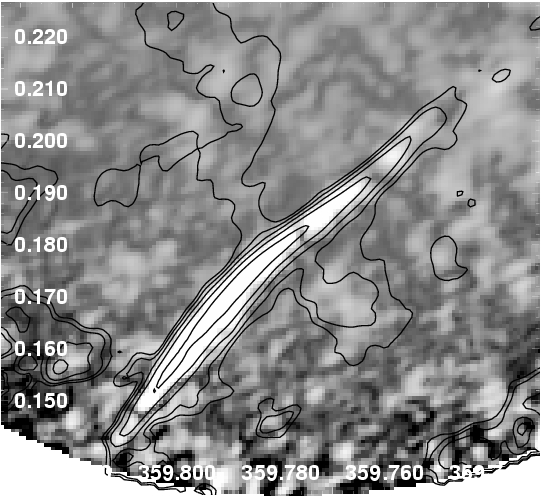}
\hfil
\includegraphics[scale=0.44,angle=270,bb=540 130 575 692]{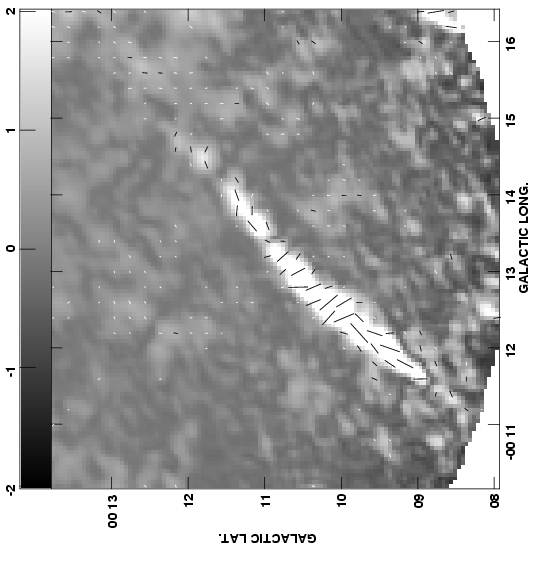}
\caption{\emph{Left}:  Gray scale shows the 6cm polarized intensity from RF-N8 (G359.79+0.17) for brightnesses ranging from -1 to 5 mJy beam$^{-1}$.  The contours show 6 cm total intensity at levels of $0.5*2^n$ mJy beam$^{-1}$, for $n=0-4$. \emph{Right}:  Image of the same region with the vectors showing the 6 cm polarization angle averaged over the 100 MHz band.  Vectors are shown for regions with polarized intensity brighter than 0.2 mJy beam$^{-1}$.  \label{n8pol}}
\end{figure}

Figure \ref{n8pol} shows that the 6cm polarized intensity from RF-N8 is broken in to several islands by depolarization canals.  Generally, the polarized brightness follows the total intensity brightness.  The maximum polarization fraction is 55\%, which is found at the peak brightness of two or three of the brightest polarized intensity islands.  Averaging over the polarized emission seen at 6 cm finds \dt$=-4\pm1$\sdeg, equivalent to \rmeff$=880\pm220$ rad m$^{-2}$.

\begin{figure}[tbp]
\includegraphics[width=6.5in]{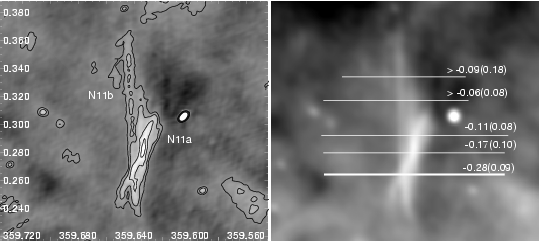}
\includegraphics[width=6.5in]{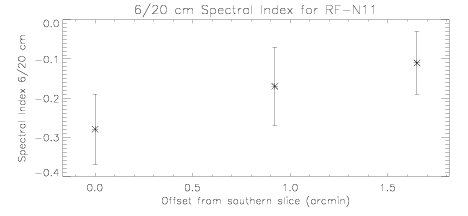}
\caption{Same as Figure \ref{n10fig}, but for the N11 radio filament.  The left image has contour levels at 0.5, 1, 2, and 3 mJy per 15\arcsec$\times$10\arcsec-beam.  The thick slice in the right image shows the southernmost slice, which is also the origin of the plot, at (l,b) = (359\ddeg640,0\ddeg264);  the spectral index is measured along a line oriented 3\sdeg East of Galactic North. \label{n11fig}}
\end{figure}

$G359.62+0.28 (RF-N11a)$ --- Figure \ref{n11fig} shows a complex of filaments near ($l$,$b$) = (359.6,+0.3), first noted in the 20 cm observations of Paper I and named ``RF-N11''.  The 6 cm observations presented here resolve the N11 filament into two sets of crossed filaments:  one, 100-mJy, kinked filament running southeast to northwest, and then one or two parallel filaments running north-south with a flux density of about 60 mJy.  Thus, in this work, the kinked, diagonal filament is referred to as N11a and the vertical filaments are called N11b.  N11a is detected at 6 and 20 cm, and the spectral index is relatively flat with values ranging from --0.28 to -0.11.  There is no significant trend for the index values to change along the filament.

\begin{figure}[tbp]
\includegraphics[scale=0.44]{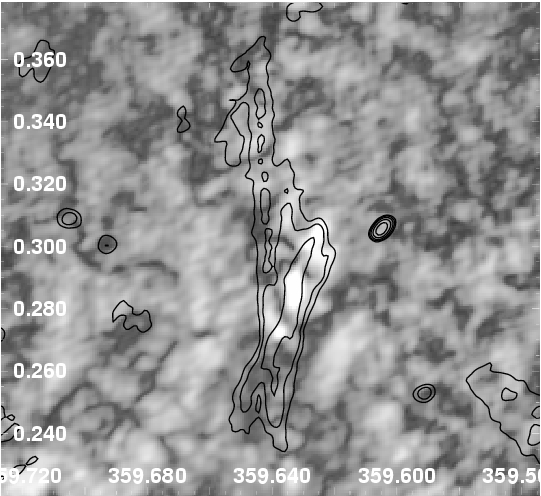}
\hfil
\includegraphics[scale=0.44,angle=270,bb=500 0 600 500]{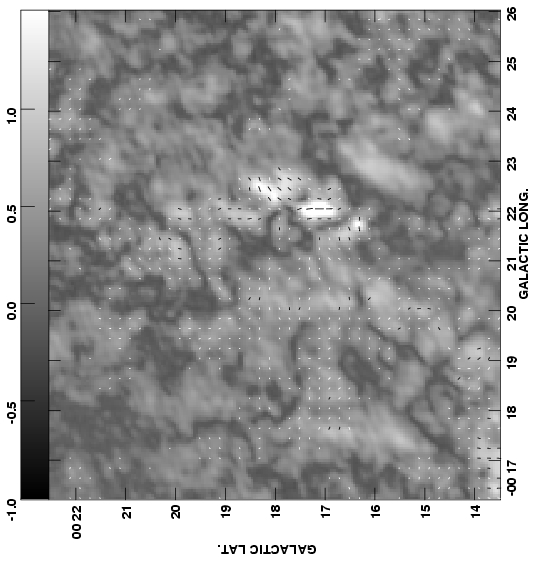}
\caption{\emph{Left}:  Gray scale shows the 6cm polarized intensity from RF-N11 (G359.62+0.28) for brightnesses ranging from -1 to 5 mJy beam$^{-1}$.  The contours show 6 cm total intensity at levels of 0.5, 1, and 2 mJy beam$^{-1}$.  \emph{Right}:  Image of the same region with the vectors showing the 6 cm polarization angle averaged over the 100 MHz band.  Vectors are shown for regions with polarized intensity brighter than 0.2 mJy beam$^{-1}$. \label{n11pol}}
\end{figure}

Figure \ref{n11pol} shows that the brightest 6 cm polarized emission from the RF-N11 complex is broken into three islands along the N11a subfilament.  This is the first detection of polarized emission from N11a, which confirms that it is a NRF.  The peak polarized brightness is 2 mJy beam$^{-1}$, where the polarization fraction is about 70\%.  As in RFs C12 and N1, the polarized emission is brighter than the total intensity near the end of the filament, leading to apparent polarization fractions larger than 100\%.  

The average \dt for the polarized emission from RF-N11a is $0\pm1$\sdeg, or \rmeff$=0\pm220$ rad m$^{-2}$.  There is weak evidence for changes in \dt across N11a, with \dt$=1\pm2$\sdeg\ (\rmeff$=-220\pm440$ rad m$^{-2}$) and $-1\pm2$\sdeg\ (\rmeff$=220\pm440$ rad m$^{-2}$) for the northern and central polarized islands, respectively.

$G359.64+0.30 (RF-N11b)$ --- The fainter, vertical filaments next to N11a in Figure \ref{n11fig} are here referred to as N11b.  N11b seems to be composed of two parallel filaments that cross the N11a filament near its midpoint.  N11b is seen in multiple VLA pointings, and thus is unlikely to be an imaging artifact.  As seen in Figure \ref{n11pol}, there is polarized emission throughout this region, but no clear morphological association with the N11b filament.  

The N11b filaments have no counterpart at 20 cm to a level of 0.2 mJy beam$^{-1}$ in the present observations nor in Paper I, which constrains the spectral index to be greater than about --0.09.  This spectral index is unusually flat for NRFs, but is more common in NRFs found in groups \citep{m96}.  The limit on the spectral index is not plotted at the bottom of Figure \ref{n11fig}, but the values are consistent with a further flattening of the spectral index in N11a with increasing Galactic latitude.

\subsubsection{Results for Other Extended Objects}
Galactic Center Lobe --- Figures \ref{6large} and \ref{20large} show that the emission from the GCL dominates the images made from the feathered data.  The brightness of the GCL appears strongest on spatial scales around 0\ddeg2, which is near the largest angular scale that the 6 cm observations are sensitive to ($\theta_{LAS}^{6 \rm{cm}}\approx0\ddeg1$).  The largest angular scale that the 20 cm observations are sensitive to is about three times larger ($\theta_{LAS}^{6 \rm{cm}}=900$\arcsec), so some of the GCL emission is visible in the 20 cm VLA-only image.  Since the GCL emission is not fully sampled by both data sets, no spectral index study can be done with the VLA data.  The radio spectral index distribution of the GCL from the GBT data are discussed in detail elsewhere \citep{gcl_all}.  Nonetheless, it is noteworthy that these first high spatial resolution observations of the GCL have constrained the spatial scale on which it emits.  Prior to these observations, single-dish observations could not determine if the GCL was a single, shell-like structure on 0\ddeg2-scales, or composed of smaller-scale filamentary structures.

\begin{figure}[tbp]
\includegraphics[width=6.5in]{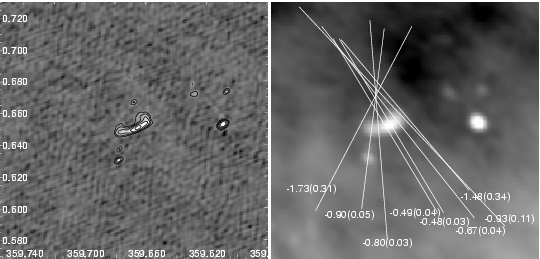}
\includegraphics[width=6.5in]{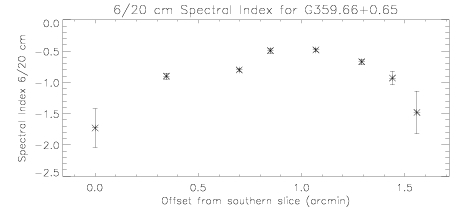}
\caption{Same as Figure \ref{n10fig}, but for G359.66+0.65. Contour levels are at 0.25, 0.5, 1, 2, and 4 mJy per 15\arcsec$\times$10\arcsec-beam.  The plot origin is at (l,b) = (359\ddeg681, 0\ddeg648) and it is oriented 295\sdeg East of Galactic North.  \label{359.66+0.65fig}}
\end{figure}

$G359.66+0.65$ --- Figure \ref{359.66+0.65fig} shows the 6 cm emission and spectral index across G359.66+0.65, an elongated, curving source roughly 1\damin5 long.  G359.66+0.65 has an integrated total intensity flux density of $29.8\pm2.0$ mJy in the 6 cm feathered image.  The spectral index distribution is shown for a series of slices across G359.66+0.65 in Figure \ref{359.66+0.65fig}.  The index shows a steady change with position along the source ranging from --1.73 in the faintest part of the lobe-like region to --0.49 in the bright core region.  The integrated spectral index measured in the feathered images is $-0.92\pm0.12$.

G359.66+0.65 has a polarized counterpart with a flux density of $23.0\pm3.3$ mJy.  Comparing the polarized and total integrated flux densities for the VLA data give a polarized fraction of about 72\% for G359.66+0.65, which is near the theoretical maximum of $\sim$74\% for synchrotron emission polarization fraction with $\alpha=-0.9$ \citep{r79}.  No polarized source is seen at 20 cm down to a level of 0.2 mJy beam$^{-1}$;  depolarization effects scale as $\lambda^2$, so they are about 11 times higher at 20 cm than at 6 cm.

The shape, location, and spectral index of this source are most consistent with FR I type galaxy.  These galaxies typically have a central bright core with symmetric, swept-back radio jets \citep{f74,ge05};  the high Galactic latitude is also most consistent with an extragalaxtic source.  However, radio filament candidates can have a similar shape, can be found at high latitudes, and may have spectral index gradients, so we cannot rule out the possibility that G359.66+0.65 is a filament.

\begin{figure}[tbp]
\includegraphics[width=6.5in]{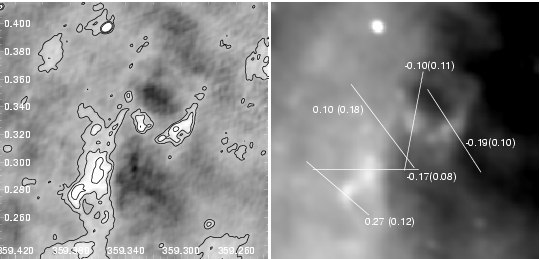}
\caption{\emph{Left}: 6 cm, VLA-only mosaic toward G359.34+0.31 with a beam size of 15\arcsec$\times$10\arcsec\ with $\theta_{PA}=70$\sdeg.  Contour levels are at 0.5, 1, and 2 mJy beam$^{-1}$.  \emph{Right}: 6 cm feathered image of the same region with spectral indices overlaid. \label{359.34+0.31fig}}
\end{figure}

$G359.34+0.31$ --- This extended source is located within the western half of the radio continuum and recombination line emission associated with the GCL \citep{u94,thesis}.  Figure \ref{359.34+0.31fig} shows the 6 and 20 cm emission, which has a wispy, complex morphology.  The spectral index for several slices across the feathered images of G359.34+0.31 are consistent with optically-thin, thermal emission.  It is difficult to interpret the structure in these observations, since the spatial scale of this emission is almost too large to be visible to these VLA 6 and 20 cm observations.  This source and its relation to the GCL is discussed in more detail elsewhere \citep{gcl_all}.

\begin{figure}[tbp]
\includegraphics[width=6.5in]{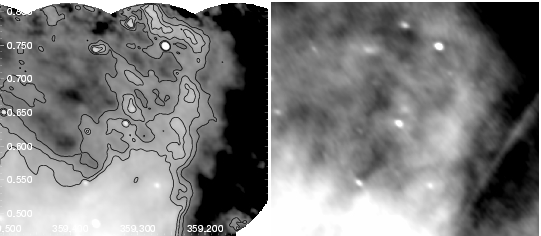}
\caption{The left and right panel show the 6 and 20 cm feathered radio continuum images of the region around G359.2+0.75, or the bend in the northwest corner of the GCL.  The contours in the left panel are at 3, 4, and 5 mJy per 18\arcsec$\times$26\arcsec-beam.  \label{359.2+0.75fig}}
\end{figure}

$G359.2+0.75$ --- At the northwest of the GCL, the radio continuum morphology bends toward higher longitudes, making a hooked shape similar to that seen in the 6 cm GBT image.  Figure \ref{359.2+0.75fig} shows the 6 and 20 cm feathered images of this structure, here referred to as G359.2+0.75.

The spectral index varies from --0.6 to +0.4 with the steepest index measured south of the bend the inverted index measured north of the bend.  The slice values are not shown in Figure \ref{359.2+0.75fig}, since the variation in $\alpha_{LC}$, including strongly inverted index values, are inconsistent with the GBT only slice analysis \citep{thesis}.  The VLA's largest resolvable angular scales of 5\arcmin\ and 15\arcmin\ at 6 and 20 cm, so emission on these size scales are more uncertain.  This uncertainty is likely to be worst near G359.2+0.75 and G359.1+0.9 (described below), since they have complex distributions of emission on scales around 5\arcmin.

\begin{figure}[tbp]
\includegraphics[width=6.5in]{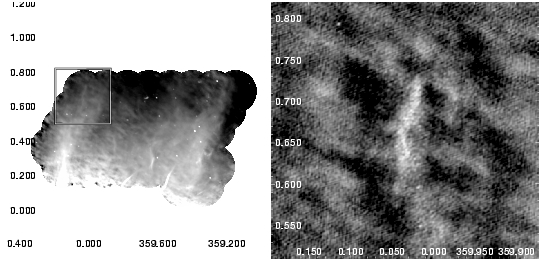}
\caption{\emph{Left}: Large-scale, 6 cm feathered image of the entire the survey map with Galactic coordinates.  The box shows the region covered by the image at right.  \emph{Right}: Image of the VLA-only data centered at G0.03+0.66 with a beam size of 9\dasec4$\times$12\dasec8 with $\theta_{PA}=66$\sdeg.  \label{0.03+0.66fig}}
\end{figure}

$G0.03+0.66$ --- Figure \ref{0.03+0.66fig} shows large- and small-scale views of G0.03+0.66, a 6\arcmin-long, wavy structure seen in 6 cm VLA data.  There is no counterpart to G0.03+0.66 in the 20 cm map, which suggests that it has a relatively flat spectral index, most consistent with thermal emission.  G0.03+0.66 is located within the GCL-East radio continuum emission ridge, similar to G359.34+0.31 in GCL-West.  In this case, however, the structure seen in the VLA data are oriented parallel to the structure seen by the GBT, so they are more likely to be associated with each other.  However, the radio continuum follows the wavy shape of the 8 $\mu$m emission seen by \emph{Spitzer} that has been identified as the ``Double Helix Nebula'' \citep{m06}.  The nature of this source and its possible association with the GCL is discussed elsewhere \citep{gcl_all}.

\begin{figure}[tbp]
\begin{center}
\includegraphics[width=0.8\textwidth]{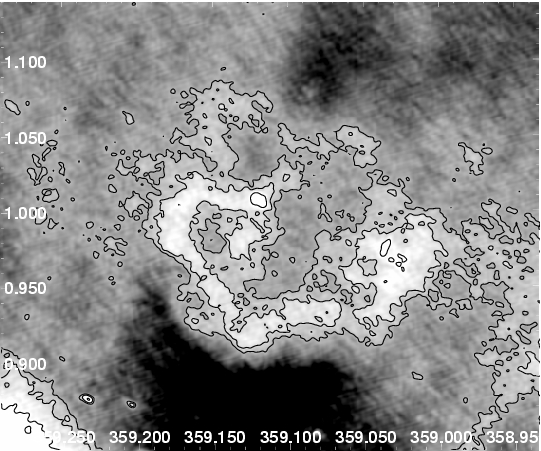}
\end{center}
\caption{Image of the 20 cm continuum emission of G359.1+0.9 seen by the VLA in Galactic coordinates. Contours are at brightness levels of 0.3, 0.6, and 1.2 mJy per 10\dasec9$\times$15\dasec9-beam \label{359.1+0.9fig}}
\end{figure}

$G359.1+0.9$ --- The 20 cm continuum emission from G359.1+0.9 is shown in Figure \ref{359.1+0.9fig}.  G359.1+0.9 is a shell-like SNR that extends 10\arcmin$\times$9\arcmin\ with an irregular ring.  This SNR is only covered by the present 20 cm survey, so no 6 cm map or spectral index analysis is done.  The 20 cm integrated flux density is 1.3$\pm$0.5 Jy.

\subsection{Compact Sources}
\label{vla_pssec}

\subsubsection{Source Detection}
The 6 and 20 cm compact source detection was performed on the VLA images optimized for sensitivity to compact emission.  This is done by removing short baselines before imaging the data giving beam sizes of 14\arcsec $\times$9\arcsec at both 6 and 20 cm (see Table \ref{image_resolutions}).  At 6 cm, sources were detected with the sfind routine of Miriad\footnote{see \url{http://www.atnf.csiro.au/computing/software/miriad/}}.  The algorithm (called ``old'' in Miriad) had a detection threshold of $5\sigma$.  To ensure accurate comparison between the 6 and 20 cm catalogs, the 6 cm integrated flux density error was calculated manually using a relation similar to AIPS.  
The SAD and JMFIT algorithms of AIPS were used on sources that sfind had difficulty with, such as obvious double sources or confused sources.  At 20 cm, all potential sources were selected by eye and detected with SAD or JMFIT.  A constant background plus elliptical Gaussian was fit to each source.  The final source list was made by cataloging all apparent sources and keeping those with $S_p>5\sigma$.  Fluxes are not corrected for bandwidth smearing, which reduces the peak brightness and slightly elongates the sources at field edges (mostly at 20 cm).
For compact source detection in polarized-continuum images, SAD and JMFIT tasks were used with a peak flux detection threshold of 3$\sigma$.

The parameters found with the sfind, SAD, and JMFIT routines were compared on several sources.  In all cases tested, the background-subtracted flux densities for all methods were equal within their quoted errors.  There is a slight trend for SAD/JMFIT to find higher peak flux uncertainties than sfind.  However, none of the several sources detected with both techniques had less than 3$\sigma$ significance in the peak brightness using either technique.  Thus, the 6 cm sources have at least 3$\sigma$ significance, while the 20 cm sources have at least $5\sigma$ significance, and both lists have flux densities that are robust to changes in the detection algorithm.

\subsubsection{Source Catalogs}
Tables \ref{ps6cm5} and \ref{ps20cm5} show the catalogs of compact sources detected on the mosaicked, high-resolution images at 6 and 20 cm, respectively.  In total, we find 93 compact sources at 6 cm and 48 sources at 20 cm.  Tables \ref{6poln} and \ref{20poln} show the results of source detection run on polarized-continuum mosaic images;  2 and 7 compact, polarized sources are found at 6 and 20 cm, respectively.  The results of source detection include the source positions, the background-subtracted peak and integrated flux density and their uncertainties, and the apparent source size (not deconvolved) and orientation.  

\begin{figure}[tbp]
\begin{center}
\includegraphics[width=\textwidth]{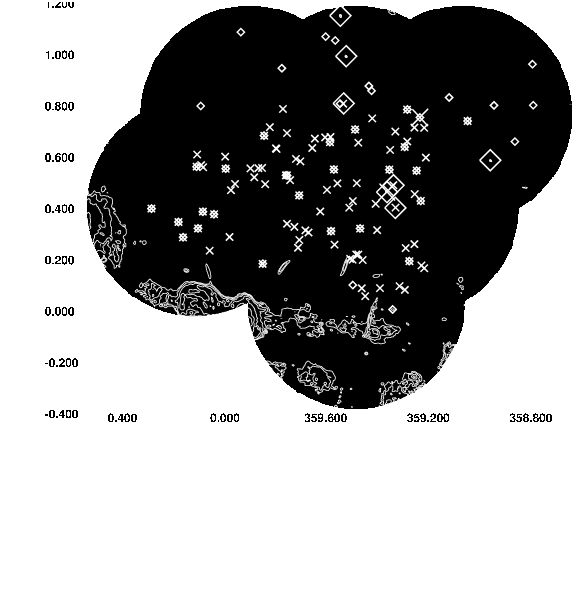}
\end{center}
\caption{The distribution of compact sources given in Tables \ref{ps6cm5}, \ref{ps20cm5}, \ref{6poln}, and \ref{20poln} are shown against the 20 cm mosaic.  Crosses and diamonds show 6 and 20 cm sources, respectively;  larger symbols show polarized compact sources.  Contour levels are at 0.02, 0.04, 0.08, 0.16, 0.32, 0.64 Jy/beam. \label{compsrc}}
\end{figure}

Figure \ref{compsrc} shows how these sources are distributed across the survey region.  In general, the distribution uniform and shows no sign of enhanced density near the Galactic plane;  this is discussed in more detail in \S \ref{vla_compact}.  About 17 of the 93 sources in the 6 cm catalog have an apparent size greater than 1.5 times the beam.  However, all these source have brightnesses less than a few mJy ($\lesssim10\sigma$ detections), so they are poorer quality fits.  At 20 cm, roughly 28 of 48 sources have an apparent size greater than 1.5 times the beam size.  The relative number of apparently extended sources is due (at least in part) to bandwidth smearing, which can be a significant fraction of the beam size in parts of the 20 cm mosaic.

Comparing the compact source catalogs, we find 33 sources that are found in both the 6 and 20 cm catalogs and 60 sources found only in the 6 cm catalog.  There are no sources found only in the 20 cm catalog in the region shared by both surveys.  Tables \ref{ps6-20} and \ref{ps6no20} list the spectral indices (or limits) for sources found in both surveys and only the 6 cm survey, respectively.  The integrated flux densities shown in Tables \ref{ps6cm5} and \ref{ps20cm5} are used to calculate the spectral indices.  Lower limits on $\alpha$ are based on the $3\sigma$ upper limit on the 20 cm nondetection.  The significance limit for matching sources in the two catalogs is $3\sigma$.

\subsubsection{Positional Accuracy}
The absolute coordinate alignment of the compact source catalogs was tested by measuring the mean offset between compact radio sources unambiguously matched with existing catalogs.  Since the source density is low, most sources with counterparts in existing surveys were unambiguous.  Sources from the present 6 and 20 cm catalogs were compared to each other and to those of \citet{n04}, Paper I, \citet{l98}, and \citet{b94}.  These catalogs were chosen for comparison because they have the highest resolutions and best sky coverage at these wavelengths.  No significant offset was found between the present 6 and 20 cm catalogs and that of \citet{l98}, which gave the smallest rms in the position differences of 0\dasec36 for the 6 cm catalog (5 sources), and 2\dasec2 for the 20 cm catalog (7 sources).  These values are the best estimates of the absolute pointing accuracy of our catalogs.  There was no significant offset between fifteen 20 cm sources matched to the catalog of (Paper I, only sources north of $b\sim0\ddeg1$).  Between the 6 and 20 cm catalogs presented here, no significant offset was found, with ($\Delta$RA,$\Delta$Dec) = (0\dasec09$\pm$1\dasec5,0\dasec01$\pm$1\dasec5) for 32 sources.  There was a $\sim1\sigma$ offset between the eleven 6 cm sources matched to the catalog of \citet{b94}, with offsets of (--0\dasec31$\pm$0\dasec52,0\dasec99$\pm$0\dasec87);  this offset is only marginally significant and is ignored.  There was a significant RA offset between both the 6 and 20 cm catalogs with the 90 cm catalog of \citet{n04}.  The mean RA difference between the 19 sources common between the present surveys catalog and the 90 cm catalog is $RA_{20,6 \rm{cm}}-RA_{90 \rm{cm}}=-4$\dasec74$\pm$2\dasec82.  This error is about twice their quoted astrometric accuracy \citep[2\dasec1;][]{n04} and has since been attributed to an error in that catalog \citep{n04}.

\subsubsection{Consistency Between 20 cm catalog and Literature}
As a test of the accuracy of the flux densities in the source catalog, sources common with the catalog of Paper I were compared.  The fifteen sources (only those north of $b=$0\ddeg1 were considered for simplicity)  common to the two catalogs are shown in Table \ref{yhc}.  The weighted mean offset for these two catalogs is $-0.18\pm0.94$ mJy.  No trend is evident in the distribution of flux density difference for these sources;  all sources are within 3 sigma of the linear trend.

The 20 cm catalog shares 7 sources with that of \citet{l98}, as shown in Table \ref{1lc}.  That work observed in the BnA configuration with a similar sensitivity as the present 20 cm survey (0.5--2 mJy beam$^{-1}$).  The present survey had a mean frequency at 1425 MHz, compared to the two bands at 1658 and 1281 MHz for \citet{l98}.  The expectation is that our flux density measurement should fall near to the mean of their two measurements.  This is true for the following sources, all of which have matching flux densities within 3$\sigma$: 359.076+0.547, 359.259+0.74, 359.073+0.735, 359.388+0.460, 359.088+0.426.  However, excluding one source, 359.388+0.460, for having a source detection problem\footnote{The source detection of \citet{l98} shows this source to be unresolved at 1658 MHz and highly resolved at 1281 MHz.  Considering that there is unlikely to be any difference in source structure over this narrow range of frequencies, this is likely an error in source detection, which makes the flux density unreliable.}, the flux densities presented here tend to be higher than that of \citet{l98}.  The average difference between this survey and that of \citet{l98} is 20 mJy, which is much larger than the typical uncertainties of about 1 mJy.  In particular, two sources, 0.305+0.394 and 359.872+0.178, are significantly brighter in the present survey than in \citet{l98}.  The observations described in \citet{l98} are less sensitive to extended sources and may resolve out some flux.  Section \ref{vla_compact} describes the possibility that extragalactic sources seen through the GC region are affected by electron scattering \citep{l98b}.

\subsubsection{Consistency Between 6 cm catalog and Literature}
As a consistency check, the flux densities of sources detected in both the present 6 cm catalog and that of \citet{b94} were compared.  \citet{b94} observed with the CnB array with a resolution of 4\arcsec\ for a typical sensitivity of 2.5mJy.  That survey observed in a hexagonal pattern with neighboring pointings spaced by 10\arcmin\ with integration times of about 90 s.

Table \ref{gpsr5} shows that, of the 13 sources detected in both surveys, the flux densities measured here are uniformly higher than that of the \citet{b94}.  The typical difference in flux densities for sources in both surveys is about 6 mJy, much larger than the $\sim$mJy-sensitivity in the two surveys.  We have considered several possible origins for this difference.  Confusion, or source detection problems do not seem significant, since the peak brightness is similar to the integrated flux density and fit residuals are small compared to the source peak brightness, as expected for a good fit.  To test for possible calibration problems, one of the four days of 6 cm data was reimaged in one field;  the fluxes in that field were similar of those in using all the data.  Both surveys used the same flux calibrator, 3C286, and list integrated flux densities corrected for primary beam attenuation.  The sources are detected over a large area ($\sim$1\sdeg), so intrinsic variability seems unlikely.  Finally, one field from the survey of \citet{b94} was calibrated and imaged, confirming that the flux density of sources in that field are equal to the flux density in their catalog.

One possible explanation for the discrepancy is the difference in integration times for the two surveys.  The shorter integration times used by \citet{b94} reduces the sensitivity of the observations and may increase confusion that would bias the measured brightness.  Another possibility is that the surveys were conducted in different configurations of the VLA and have sensitivities to emission on different angular scales.  \citet{b94} imaged their data while excluding the shortest ($<90$ m) baselines, which makes it insensitive to emission on angular scales larger than $\sim$120\arcsec.  The present survey was observed in a more compact configuration and included baselines as short as 30 m, so it is sensitive to emission on angular scales up to $\sim$400\arcsec.

\subsubsection{Spectral Index Distribution Between 6, 20, and 90 cm}
\begin{figure}[tbp]
\includegraphics[width=\textwidth]{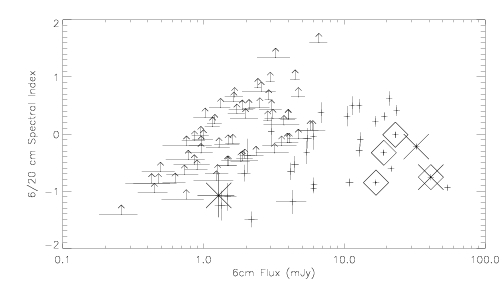}
\caption{A flux-spectral index diagram for compact sources detected at 6 and 20 cm.  Lower limits to the spectral index are shown with arrows.  The two sources polarized at 6 cm are shown with crosses and the four 6/20 cm sources with 20 cm polarization are shown with diamonds. \label{cmd206}}
\end{figure}

The distribution of spectral index of compact sources is shown in Figure \ref{cmd206}.  Sources with polarization are indicated with larger symbols.  All 93 sources presented in this paper are detected at least at 6 cm, so they either have a measurement or a limit on $\alpha_{LC}$ and appear in the figure.  The brightest sources ($S_{6{\rm{cm}}}>10$ mJy) tend to have nonthermal spectral indices, but fainter sources have a typical index (or lower limit) near 0.  These sources may be optically-thick, Galactic \hii\ regions and are described in more detail in \S \ref{vla_compact}.

\begin{figure}[tbp]
\includegraphics[width=\textwidth]{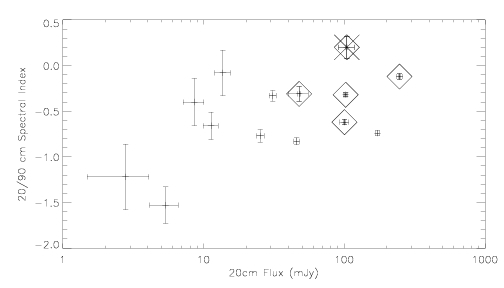}
\caption{A flux-spectral index diagram for the fourteen 20 cm sources with 90 cm counterparts in \citet{n04}. The five, 20/90 cm sources with 20 cm polarization are shown with diamonds and the one source polarized at 6 cm is shown with a cross. \label{cmd9020}}
\end{figure}

Figure \ref{cmd9020} shows the distribution of $\alpha_{20/90}$ as a function of 20 cm flux density for the fourteen 20 cm sources detected at 90 cm by \citet{n04}.  Most of the sources have 20/90 cm spectral indices between --0.3 to --0.8 for flux densities greater than 10 mJy.  This range is flatter than the range of --0.6 to --1.2 observed in other 90 cm-limited samples of spectral index \citep[for a similar extrapolated flux range][]{g04,n04}.  This range of spectral indices is typically dominated by extragalactic sources like radio galaxies, as has been argued in the field of M31 \citep{g04}.  Section \ref{hyper} discusses the possibility that a scattering screen affects the source detection.

Figures \ref{cmd206} and \ref{cmd9020} show that all polarized, compact sources are among the brightest total intensity sources.  Both 6 cm polarized sources have 20 cm total intensity counterparts, three sources are only polarized at 20 cm, and one source is polarized at both 20 cm and 6 cm.  The source polarized at 6 and 20 cm is G359.388+0.460.  \citet{r05} find the source has 3\% polarization fraction at 8.5 GHz, which is similar to that observed here at 5 GHz;  the polarization fractions increase at longer wavelengths with $3.6\pm0.8$\% at 6 cm (5 GHz) and $21.8\pm2.7$\% at 20 cm.f

\begin{figure}[tbp]
\includegraphics[width=\textwidth]{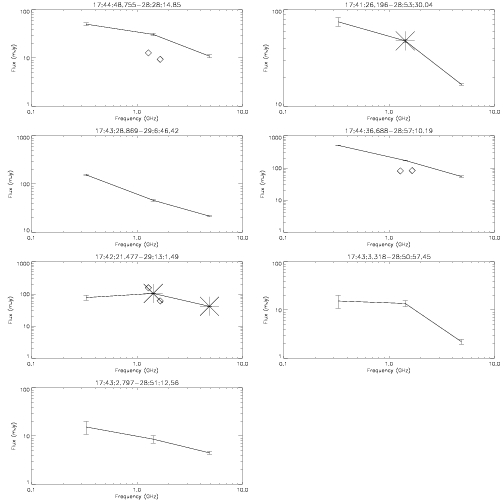}
\caption{Plots of flux density distribution for seven sources that have 90, 20, and 6 cm flux densities measured. Small diamonds show fluxes from \citet{l98}, where available, and large stars show sources that have linear polarization here. 
\label{spectra90}}
\end{figure}

Comparing our 6 and 20 cm catalogs to that of \citet{n04} finds 14 sources in common, as shown in Table \ref{gcps}.  All 6 cm sources with 90 cm counterparts also have 20 cm counterparts here.  Note that the resolution of the 90 cm survey (7\arcsec$\times$12\arcsec) is almost identical to that of the surveys presented here, so the spectral indices should be fairly accurate.  Figure \ref{spectra90} shows a more detailed look at the flux densities for the seven sources that have counterparts at 90, 20, and 6 cm.  Most of them (5/7) have spectra steepening with increasing frequency, while one has a nearly constant spectral index, and one has a flattening spectrum.

Of the pulsars and pulsar candidates listed in \citet{n04}, only one is detected in the present surveys, mostly due to our smaller spatial coverage.  This source is called 174020--291344 \citep[a.k.a. ``359.145+0.826'';][]{n04}.  The pulsar candidate has an estimated spectral index of $\alpha_{20/90}=-1.4$ by comparing 20 and 90 cm fluxes from \citet{c98} and \citet{n04}, respectively.  However, the 20 cm flux density of 174020--291344 measured by the present survey implies a spectral index $\alpha_{20/90}=-0.6\pm0.15$.  This value is significantly flatter than measured previously and is flatter than that expected from typical radio pulsars \citep[$\alpha_{20/90}\lesssim-1.0$;][]{ma05}.

\section{Discussion}
\label{vla_dis}
\subsection{Radio Filaments}
\subsubsection{General Properties}
In total 15 radio filaments were detected in the present survey.  Eight of the filaments were detected in images of linearly polarized emission, including three for the first time, thus confirming that they are NRFs.  In general, the brightest filaments in total intensity are detected in polarized emission.  The morphologies filaments with polarized emission is also similar to that of the filaments with no detected polarized emission \citep[][Paper I]{n04}.  These facts imply that number of confirmed NRFs is limited mostly by the sensitivity of polarized continuum observations.

This survey found that most of the filaments are more complex than isolated, single threads.  Of the fifteen detected here, six are grouped into pairs and four are composed of multiple parts.  In fact, many high-resolution studies of filaments generally find complex morphologies \citep[e.g.,][Paper I]{l95}.  In general, there are three ways that filaments are grouped:  (1) groups of parallel filaments, (2) single filaments that split at their ends, and (3) crossed filaments.  

The polarized emission from the NRFs generally correlates with the total intensity.  In general, filaments observed here have a peak polarization fraction of at least 30\% at 6 cm.  The theoretical prediction for synchrotron polarization fraction is 75\% for a spectral index of --1 \citep[assuming $S_\nu\propto\nu^{\alpha}$;][]{r79}, so the depolarization toward an NRF is typically about 60\% at 6 cm.  Since bandwidth depolarization isn't expected to be strong, this depolarization must be caused by either beam depolarization or depth depolarization \citep{u03,h04}.

\subsubsection{Unusual Filaments}
The filament complex N11 (G359.64+0.30) has an unusual 6/20cm spectral index.  Most filaments have 6/20 cm spectral indices around --0.7 to --0.9 \citep{l95,l99a,l99}, but few have flat spectral indices (previous known examples are G0.2+0.0 and G359.96+0.09; \citet{y86,a91}).  This similarity is surprising because these examples have different brightnesses (G0.2+0.0 is roughly a factor of 100 brighter at 6 cm) and locations (found in and out of the Galactic plane).  Models for the generation of NRFs should account for the similar spectral indices of these complexes, despite their differences.  Furthermore, the flat spectrum of some filaments suggests that 6 cm and shorter wavelength observations may find new filaments.

RF-C16 is detected at 6 cm at Galactic latitudes up to 0\ddeg75, making it the highest-latitude filament detected.  RF-C16 and G358.85+0.47 \citep[also known as ``the Pelican''][]{l99a} are the furthest filaments from the GC with an offset of about 1\ddeg4 or 200 pc, assuming a distance of 8 kpc.  The total projected area covered by GC filaments and candidates ranges from $l=358.5$ to +0.5 and $b=-0.6$ to +0.75 (Paper I), or equivalently a roughly circular region of radius 0\ddeg9, centered near (l,b)=(359.6,0).  Assuming these filaments fill an equivalent 3-d space, the total volume occupied by filaments and candidates is roughly $10^7$ pc$^{-3}$.  If filaments require a minimum magnetic field strength to form, this volume gives a lower limit to the total magnetic energy in the GC region of $E_B > U_B * V = B^2/8\pi * V$.  Recent estimates for the large-scale magnetic field strength in the central few degrees is $\sim10\mu$G \citep{la05}, which gives $E_B > 1.1*10^{51}(B/10 \mu\rm{G})^2$ ergs.

\subsubsection{Spatial Gradients in Spectral Index}
This survey uniformly studied many filaments at 6 and 20 cm, so we have several new measurements of the 6/20 cm spectral index.  The typical spectral index for these filaments range from --0.5 to --0.9, but two filaments (N1 and N11b) have spectral indices near 0.  Five radio filaments (C7, C12, N1, N2, and N8) have possible spatial gradients in their 6/20 cm spectral indices on arcminute scales, with index changes ranging from 0.4 to 0.7.  Two filaments (C7 and C12) show rapid changes in the spectral index, changing from $\sim$--0.7 to $\sim$--1.1 as one moves across the peak brightness of the filament.  Perhaps coincidentally, three filaments with significant inclinations east of Galactic north (C7, C12, and N8) have steeper indices toward their southeast ends.  However, two filaments oriented roughly perpendicular to the plane (N1 and N2) tend to steepen toward the north.

There are a few observational effects that could create a spatial gradient in a spectral index, although they cannot explain gradients seen in all of the filaments.  Most biases and artifacts are strongest at the edge of each field, where bandwidth smearing is strongest and primary beam corrections are largest.  However, the survey pointings were organized to reduce sensitivity variations to at most a factor of two between pointings.  In the case of two filaments (C7 and C12), the spectral index change does not occur across a field edge.  Bandwidth smearing toward filaments at 20 cm can be estimated from Figures 18-1 and 18-2 of \citet{b99} and accounting for the relative orientation of the filament to the radial distance from the nearest phase center.  Three filaments (N1, N2, and C12) are oriented radially to their nearest 20 cm field phase center, so smearing does not noticably reduce their peak brightness.  Two other filaments (N8 and N10) are oriented perpendicular to the radial line from the nearest 20 cm field and their indices may be biased (see \S\ \ref{nrfsec}).  Comparing slice and integrated analysis of compact sources near the edge of the 20 cm primary beam covering Filaments C3, C6, C7, C11, C12, and N11 shows that they are not significantly affected by bandwidth smearing.

There have been several previous studies of the spatial and spectral changes in the spectral index of NRFs \citep{l00,l01,so92,l95,l99}.  Most previous studies of NRFs have found that the spectral index does not change across their lengths, but those sources are nearer to the Galactic plane than the present work \citep{l99,l00,l95}.  If the observed spatial gradients in the indices are real, then these gradients seem to be more common away from the Galactic plane.  This would be consistent with the steepening of the spectral index of G0.2+0.0 (the Radio Arc) at high Galactic latitidues \citep{p92} and spatial gradient in the high-latitude ($b\sim0$\ddeg4) filament N10 \citep{l01}.

One interpretation for spatial gradients in a synchrotron spectral index is to attribute the changes to the relative ``age'' of the plasma \citep[e.g.,][]{j73}.  Electrons with higher energies emit power more efficiently and cool more rapidly through a process called synchrotron cooling.  Synchrotron cooling preferentially reduces the flux at higher frequencies, breaking the spectrum at a frequency $\nu_b\propto B^{-3}t^{-2}$ \citep{b00}.  Several filaments have been observed with steeper spectral indices at higher frequencies \citep{l99,l01,y05}, which is commonly attributed to synchrotron cooling of filament as a whole.  This model can also explain the spectral index spatial gradients if electrons propagate from a single acceleration region;  the steeper spectral indices are interpreted as being further from the acceleration region.  For a 1 mG magnetic field and a spectral break between 6 and 20 cm, synchrotron aging implies an age of $2\times10^4$ yr, which is consistent with estimates of the electron diffuse velocity in an NRF \citep{l01}.

Alternatively, \citet{l01} note that the spatial gradient in the filament spectral index may be a consequence of how a curved electron energy distribution emits in a spatially changing magnetic field.  The observed frequency of synchrotron emission depends on the electron energy and magnetic field as $\nu_{\rm{syn}}=c B^2 E_{el}$, so electrons observed at a given frequency will have different implied energies if the magnetic field strength changes.  A spatial gradient in the spectral index could indicate a weaker magnetic field, as follows:  $\delta\alpha=a_c log(B_1/B_2)$, where $a_c$ is the curvature in the electron energy distribution, and $B_1$ and $B_2$ are the magnetic field strengths across the region where the spectral index is observed to change \citep{b00}.  The observed spectral index changes in filaments show a trend to steepen from $\alpha\sim-0.5$ to $-1.5$ for frequencies from 0.3 to 5 GHz \citep{l99,l01}.  Using the longest-wavelength spectral index as an reference, a change of $\Delta\alpha=-1$ from $\nu=0.3$ to 5 GHz corresponds to a spectral curvature $a_c\approx1.5$.  Assuming this curvature for the electron energy distribution and a spectral index spatial gradient across the filaments of $\delta\alpha\sim0.5$ (similar to that seen here), the magnetic field strength across a filament falls by about a factor of two.  Interestingly, in the case of two filaments (C7 and C12) there is a correlation between brightness changes and index changes;  it is possible that these changes are tied to changes in the structure of the NRF.  Unfortunately, very few filaments have clear detections of spectral curvature, so this model has not been tested outside of the present work and that of \citet{l01}.

\subsection{Nature of Compact Sources}
\label{vla_compact}
\subsubsection{Spectral Index and Polarization Properties}
Figure \ref{cmd206} shows that the 6/20 cm spectral index distribution follows the roughly bimodal distribution seen in other surveys \citep{b94}.  Flat or positive (``inverted'') spectrum sources, with $\alpha_{LC}\gtrsim0.0$ are likely to be \hii\ regions or planetary nebulae \citep{g05b,i81}.  Positive spectral indices are more likely to be optically-thick, thermal emission.  A flat radio spectral index is also observed toward pulsar wind nebulae \citep{h87,ge05}.  Sources with $\alpha\approx-0.6$ are more likely to be FR I and FR II radio galaxies or supernova remnants \citep{f74,g04,ge05}.  

An optically-thin \hii\ region in our Galaxy (within 20 kpc) excited by a B0-type star or earlier would have $S_{\rm{6 cm}}>26$ mJy and $\alpha_{LC}=-0.1$ \citep{g05a}.  These values of flux density and spectral index correspond to a region of the plot in Figure \ref{cmd206} that is basically empty, so such objects are not expected in this survey.  However, there are about two dozen sources with flux densities between 1 and 10 mJy and positive spectral indices, which would be expected for \hii\ regions that are optically thick at 6 cm.  Alternatively, planetary nebulae are expected to have a similar distribution of fluxes and spectral indices as \hii\ regions \citep{i81}, although they are much less numerous in the Galaxy \citep{g05a}.

Only two sources presented here have spectral indices approaching values commonly associated with pulsars, with $\alpha\lesssim-1.5$ \citep{lo95};  these are the sources at (17:40:33,--28:44:05) (between 90 and 20 cm; Table \ref{gcps}) and at (17:43:03,--28:50:57) (between 20 and 6 cm; Table \ref{ps6-20}).  Taking a higher index threshold of $\alpha<-1.0$, includes five more candidates, at (17:42:07,--29:18:51), (17:40:55,--29:06:53), (17:42:10,--29:20:53), (17:43:20,--28:36:33) (at 20 and 6 cm) and (17:41:49,--28:43:20) (at 90 and 20 cm).  Steep radio continuum spectra can also be found in high-$z$ radio galaxies and relic radio galaxies, where synchrotron aging steepens the electron energy (and hence radio emission) spectrum \citep{k00,ge05}.  H{\small I} absorption observations may help localize the source location and constrain its nature.


There are significantly more compact sources polarized at 20 cm, (7, 3$\sigma$ detections) as compared to 6 cm (2 detections).  Both surveys have similar sensitivity to polarized emission, with typical errors in the integrated, polarized, compact flux density of 0.3 mJy beam$^{-1}$.  For compact sources, depolarization is complete when Faraday rotation rotates the polarization angle by 180\sdeg across the band (called ``bandwidth depolarization'').  Since depolarization effects increase as $\lambda^2$ and most sources have spectral indices flatter than --2, the naive expectation is that more polarized sources would be found at short wavelengths (for a given sensitivity).  The detection of more 20 cm polarized sources suggests that the population of polarized sources must have a flux density that increases more rapidly than $\lambda^2$.  This suggests that the sources have fairly steep, nonthermal spectra and/or an intrinsically increasing polarization fraction with increasing wavelength.  As mentioned above, sources with radio spectral indices steeper than --1.3 are often identified with ``relic'' radio galaxies, which have radio lobes with steep spectra due to synchrotron cooling of high radio frequency emission \citep{ge05}.  Pulsars are also known to have steep radio spectra and emit polarized radio continuum, so they may contribute to the compact polarized sources observed in this survey.

\subsubsection{Spatial Distribution}
There is no evidence for variation in the number density of sources (thermal or nonthermal) over the range of Galactic latitudes covered in this survey, from $l=0$\ddeg1--0\ddeg8.  Considering that the scale height of Galactic \hii\ regions is 24\arcmin--28\arcmin\ \citep{g05b}, one might expect variation in the number density of sources with Galactic latitude, particularly flat- or inverted-spectrum (likely Galactic) sources.  The number density of thermal, compact sources is about 37 per square degree, which is about twice the source density of all sources away from the Galactic plane \citep[see Fig. 2 of][]{b94}.  Therefore, many of the sources observed in the present survey are Galactic.  The uniform number density of these Galactic sources is more likely caused by changes in the survey's sensitivity with Galactic latitude.  The 6 cm noise level near the Galactic plane is about four times higher than the typical 6 cm noise level because of confusion with sidelobes from Sgr A* and larger contribution of sky emission to the system temperature.

\subsubsection{Hyperstrong Scattering Affecting Flux Measurements?}
\label{hyper}
Comparing the flux densities of sources between various catalogs shows a trend for higher fluxes in the present survey.  Since this is a statistical effect, it is unlikely to be caused by intrinsic variability.  We can also rule out another common effect at GHz frequencies called refractive interstellar scintillation, which is caused by a turbulent motion of electron inhomogeneities \citep{co98,ge05}.  The scintillation should only be visible when observed with bandwidths less than the correlation frequency, which is generally much smaller than our 50 MHz bandwidth \citep{l71}.  

Alternatively, the scattering of radiation by the ISM can also affect the apparent size of sources observed through the GC region.  Multiwavelength observations of Sgr A* have seen the $\nu^{-2}$ size dependence expected from strong scattering by variations in the electron density along the propagation path \citep[e.g.,][]{d76}.  \citet{l98b} used the angular broadening of sources seen through the GC region to find a region of ``hyperstrong'' scattering in the central 1--2\sdeg of the GC.  Extragalactic radio sources seen through the central degree of the Galaxy are expected to be broadened by as much as 500\arcsec\ at 20 cm \citep{v92,l98,l98b}, such that interferometers would not be sensitive to them.  Since this is basically a geometric effect, extragalactic sources suffer proportionately to their distance.  This effect has been shown to bias against the detection of 20 and 90 cm compact sources in the central degree of the GC \citep{l98b,n04}.  This ``scattering bias'' can also explain why the typical spectral index of sources detected at 90 and 20 cm in the central degree of the GC region is significantly flatter than those found toward M31 and over a larger portion of the GC region.  The 90 cm observations of the GC region are not sensitive to emission larger than $\sim$540\arcsec, so most sources detected at 90 cm toward the scattering screen are located inside the galaxy \citep{n04}.  If a scattering bias is affecting this data, then sources most discrepant should be extragalactic.

Hyperstrong scattering may also alter the fluxes of our compact source catalog.  In \S\ \ref{vla_pssec}, it was discussed how the 6 and 20 cm flux densities are generally higher than observed previously, in observations less sensitive to large-scale emission.  The data used for the compact source detection is sensitive to emission at the largest angular scale of about 240\arcsec, while \citet{b94} and \citet{l98} were sensitive to angular scales up to 120\arcsec.  Meanwhile, the interferometric observations of Paper I have multiple configurations and are sensitive to emission up to $\sim500$\arcsec, which could explain the consistency between that and the present work.  If these flux densities are discrepant because of scattering, the wavelength dependence of the scattering suggests that the spectral index measurements may be positively biased, despite the fact that the observations have similar sensitivies.  However, more work is needed to show that scattering can significantly bias the observed flux densities.

\section{Conclusions}
\label{vla_con}
This paper presents the results of 6 and 20 cm, VLA radio continuum surveys of a roughly 1 square degree near the Galactic center.  The observation was designed to have similar sensitivies at 6 and 20 cm to facilitate several spectral index studies.  This feature of the survey allowed us to make new observations of the properties of the population of NRFs in the central degrees of the Galaxy.  In total, three new NRFs were discovered by detecting their linearly polarized emission.  Generally, the brightest filaments are found to be polarized, which suggests that the number of confirmed filaments is mostly limited by the sensitivity of the data.  Noteworthy filaments include RF-C16, which extends the region occupied by the NRFs up to $b=+0.75$ and a radial distance of $\sim$200 pc from Sgr A*.  These observations also revealed an unusual, twisting NRF complex at 6 cm, providing us with a rare example of a flat-spectrum NRF.  

This work also finds evidence that some of the filaments observed in this survey have spatial gradients in their 6/20 cm spectral indices.  After considering observational biases, we conclude that some of these spatial gradients are intrinsic to the filaments.  Possible models for these changes consider the effects of synchrotron cooling and changing magnetic field strengths.  If the spatial gradients are caused by changing magnetic field strengths, they are consistent with magnetic fields changing by roughly a factor of two over the filaments.  However, the present observations cannot exclude the synchrotron aging model for these gradients.  Observations of these spectral index gradients over a wider range of frequencies would better constrain the models.

Comparing the compact source fluxes between this and previous surveys suggests that electron scattering decreases the apparent brightness of some sources, especially for compact interferometers.  If this is true, it would introduce a new bias in spectral index measurements, particularly for compact extragalactic sources.

\begin{acknowledgements}
We would like to thank Ron Maddalena and Doug Roberts for contributions to the planning of these observations.  We also thank the referee, Cornelia Lang, for thorough and very helpful comments.
\end{acknowledgements}

{\it Facilities:} \facility{VLA ()}

\clearpage

\begin{deluxetable}{lcc}
\tablecaption{Parameters for VLA Continuum Observations \label{vla_params}}
\tablewidth{0pt}
\tablehead{
\colhead{Parameter} & \colhead{6 cm} & \colhead{20 cm} \\
}
\startdata
Dates & June 2004 & January--August 2004 \\
Frequency & 4.85 GHz & 1.4 GHz \\
Configuration & DnC & CnB, DnC, D \\
Integration time per pointing & $\sim$30 min & $\sim$110 min \\
Typical beam size & 12\arcsec$\times$8\arcsec & 11\arcsec$\times$8\arcsec \\
Survey area & 0.8 sq. degrees & 2 sq. degrees \\
Bandwidth & 2$\times$50 MHz & 2$\times$50 MHz \\
{\it uv} sensitivity range & 0.1 to 30 k$\lambda$ & 0.4 to 30 k$\lambda$ \\
Theoretical sensitivity & 0.03 mJy beam$^{-1}$ & 0.02 mJy beam$^{-1}$ \\
Actual sensitivity & 0.08--0.5 mJy beam $^{-1}$ & 0.2--1.7 mJy beam$^{-1}$ \\
\enddata

\end{deluxetable}

\begin{deluxetable}{lccc}
\tablecaption{Resolutions of Images in this Work\label{image_resolutions}}
\tablewidth{0pt}
\tablehead{
\colhead{Wavelength} & \colhead{Stokes Parameters} & \colhead{Resolution} & \colhead{Purpose} \\
}
\startdata
6 cm & I & $\sim$12\arcsec\ $\times$8\arcsec & Viewing fine structure \\
20 cm & I & $\sim$11\arcsec\ $\times$8\arcsec &  Viewing fine structure \\
6 and 20 cm & I & 14\arcsec\ $\times$9\arcsec & Mosaicking for compact source detection \\
6 and 20 cm & I & $\sim$26\arcsec\ $\times$18\arcsec & Feathering for extended source analysis \\
6 cm & Q, U & $\sim$16\arcsec\ $\times13$\arcsec & Polarized intensity and rotation measure analysis \\
20 cm & Q, U & $\sim11$\arcsec\ $\times$8\arcsec &  Polarized intensity and rotation measure analysis \\
\enddata

\end{deluxetable}

\begin{deluxetable}{cccccccc}
\tablecaption{Radio Filaments Detected at 6 cm \label{nrftab}}
\tablewidth{0pt}
\tabletypesize{\scriptsize}
\tablehead{
\colhead{Identifier} & \colhead{} & \colhead{} & \colhead{Length} & \colhead{$S_\nu^p$\tablenotemark{b}} & \colhead{$S_\nu^I$\tablenotemark{b}} & \colhead{} & \colhead{}\\
\colhead{($l$,$b$)} & \colhead{Name\tablenotemark{a}} & \colhead{Common Name} & \colhead{(arcmin)} & \colhead{mJy beam$^{-1}$} & \colhead{mJy} & \colhead{$\alpha_{LC}^{slice}$} & \colhead{Pol'd?}\\
}
\startdata
G359.45--0.06 & C1 &    &    &                &                & & N \\
G359.54+0.18  & C3 & Ripple & 11 &  18.0$\pm$0.5  & 486.4$\pm$61.7 & --0.5 to --0.8 & Y \\
G359.44+0.14  & C6 &    & 2  &   1.5$\pm$0.1  &   6.9$\pm$1.4  & & N \\
G359.42+0.13  & C7 &    & 3  &   7.0$\pm$0.5  &  28.2$\pm$8.3  & --0.7 to --1.1 & N \\
G359.37+0.11  & C11&    & 2  &   1.0$\pm$0.02 &   6.3$\pm$0.2  & & N \\
G359.36+0.10  & C12&    & 4  &   4.0$\pm$0.2  &  24.8$\pm$4.5  & --0.5 to --1.8 & Y\tablenotemark{c} \\
G359.21+0.54  &C16&    &$\sim10$&0.2$\pm$0.09 &   0.6$\pm$6.5  & $<-1.2$ & N \\
G0.15+0.23    & N1 & Radio Arc &    &                &                & +0.2 to --0.5 & Y\tablenotemark{c} \\
G0.08+0.15    & N2 & Northern Thread &    &                &                & --0.8 to --1.2 & Y \\
G359.96+0.09  & N5 & Southern Thread &    &                &                & & Y \\
G359.79+0.17  & N8 &    & 9  &   11.4$\pm$0.5 & 226.7$\pm$24.8 & --0.9 to --1.3 & Y \\
G359.85+0.39  & N10&    & 5  &    0.9$\pm$0.1 &  42.9$\pm$9.0  & --0.6 to --1.5 & Y \\
G359.62+0.28  & N11a&    & 4 &    4.0$\pm$0.1 & 104.8$\pm$6.2  & --0.1 to --0.3 & Y\tablenotemark{c} \\
G359.64+0.30  & N11b&    & 8 &    2.5$\pm$0.2 &  62.0$\pm$17.8 & $>-0.1$ & N \\
G359.71+0.40  & N12&    &$\sim3$& 0.4$\pm$0.06&   2.7$\pm$1.7  & & N \\
\enddata
\tablenotetext{a}{Name from YHC04}
\tablenotetext{b}{Peak and integrated flux density measured only for filaments that are completely covered by the 6 cm survey.}
\tablenotetext{c}{First confirmation that this filament is an NRF.}

\end{deluxetable}

\begin{deluxetable}{ccccccccc}
\tablecaption{Compact Sources in 6 cm Survey \label{ps6cm5}}
\tablewidth{0pt}
\tablehead{
\colhead{RA} & \colhead{Dec} & \colhead{$S_p$} & \colhead{$\sigma_{S_p}$} & \colhead{$S_i$} & \colhead{$\sigma_{S_i}$} & \colhead{$\theta_M$} & \colhead{$\theta_m$} & \colhead{$\theta_{PA}$ \tablenotemark{a}} \\  
\colhead{(J2000)} & \colhead{(J2000)} & \colhead{mJy beam$^{-1}$} & \colhead{mJy beam$^{-1}$} & \colhead{mJy} & \colhead{mJy} & \colhead{arcsec} & \colhead{arcsec} & \colhead{degrees} \\
}
\startdata
17:42:07.92 & -29:18:51.9 & 0.71 & 0.14 & 1.28 & 0.37 & 16.4 & 13.9 & 9.9 \\
17:44:48.78 & -28:28:13.2 & 9.24 & 0.36 & 10.91 & 0.70 & 14.8 & 10.4 & 54.9 \\
17:44:45.92 & -28:35:15.2 & 10.66 & 0.17 & 12.70 & 0.34 & 15.1 & 10.1 & 69.1 \\
17:44:57.38 & -28:38:02.3 & 8.09 & 0.29 & 12.75 & 0.69 & 17.1 & 12.0 & 71.6 \\
17:43:45.58 & -28:32:05.9 & 10.30 & 0.14 & 11.39 & 0.26 & 14.7 & 9.5 & 70.7 \\
17:43:34.26 & -28:30:42.6 & 0.88 & 0.11 & 1.47 & 0.24 & 19.4 & 9.4 & 62.1 \\
17:44:40.87 & -28:39:56.2 & 16.34 & 0.27 & 19.18 & 0.54 & 14.8 & 10.3 & 67.1 \\
17:43:41.63 & -28:32:47.9 & 0.85 & 0.11 & 1.26 & 0.24 & 17.1 & 10.9 & 61.1 \\
17:44:22.88 & -28:38:51.2 & 6.23 & 0.21 & 12.98 & 0.61 & 19.8 & 13.6 & 68.4 \\
17:43:42.47 & -28:33:28.1 & 3.62 & 0.09 & 4.06 & 0.17 & 14.6 & 9.9 & 67.7 \\
17:44:54.55 & -28:44:59.1 & 2.28 & 0.39 & 3.19 & 0.84 & 16.7 & 10.4 & 67.9 \\
17:44:18.97 & -28:41:24.3 & 6.95 & 0.14 & 6.83 & 0.24 & 13.9 & 8.9 & 67.8 \\
17:43:20.50 & -28:36:33.4 & 1.35 & 0.08 & 1.35 & 0.15 & 13.9 & 9.1 & 72.0 \\
17:43:31.22 & -28:38:07.7 & 5.67 & 0.09 & 6.08 & 0.16 & 14.5 & 9.3 & 65.1 \\
17:44:31.06 & -28:47:16.4 & 2.53 & 0.41 & 5.97 & 1.35 & 19.6 & 16.4 & 75.9 \\
17:43:47.70 & -28:41:49.4 & 0.84 & 0.19 & 2.57 & 0.69 & 25.2 & 14.6 & 56.6 \\
17:43:39.77 & -28:41:51.7 & 0.74 & 0.11 & 0.73 & 0.19 & 13.4 & 9.7 & 79.0 \\
17:43:17.04 & -28:43:01.7 & 2.08 & 0.09 & 2.85 & 0.20 & 15.7 & 11.6 & 70.4 \\
17:43:22.96 & -28:44:52.3 & 0.84 & 0.08 & 0.63 & 0.12 & 12.0 & 7.9 & 71.2 \\
17:43:12.35 & -28:44:33.7 & 0.48 & 0.09 & 0.43 & 0.15 & 12.6 & 9.4 & 72.7 \\
17:43:10.12 & -28:45:17.3 & 0.43 & 0.06 & 0.26 & 0.08 & 10.9 & 6.9 & 64.1 \\
17:44:37.05 & -28:57:09.5 & 48.66 & 1.29 & 54.08 & 2.36 & 15.0 & 9.1 & 68.8 \\
17:42:39.65 & -28:41:44.5 & 0.89 & 0.07 & 1.95 & 0.20 & 21.6 & 11.7 & 57.9 \\
17:43:23.12 & -28:47:54.0 & 0.56 & 0.10 & 0.86 & 0.26 & 16.7 & 12.1 & 73.3 \\
17:42:28.74 & -28:41:48.9 & 0.69 & 0.08 & 0.96 & 0.18 & 15.9 & 11.6 & 70.2 \\
17:42:44.73 & -28:45:48.7 & 0.83 & 0.06 & 1.20 & 0.14 & 15.8 & 12.2 & 91.0 \\
17:42:44.52 & -28:45:37.7 & 0.61 & 0.06 & 0.96 & 0.13 & 18.8 & 8.8 & 92.0 \\
17:42:04.91 & -28:42:10.4 & 1.49 & 0.18 & 1.88 & 0.37 & 14.9 & 11.3 & 79.7 \\
17:43:02.75 & -28:51:09.2 & 2.89 & 0.13 & 4.46 & 0.29 & 17.7 & 10.7 & 48.0 \\
17:43:03.46 & -28:50:55.4 & 2.10 & 0.13 & 2.18 & 0.23 & 14.5 & 8.8 & 78.0 \\
17:43:46.90 & -28:57:05.4 & 0.97 & 0.10 & 0.90 & 0.17 & 13.4 & 8.7 & 72.6 \\
17:42:24.48 & -28:45:57.6 & 0.44 & 0.06 & 0.45 & 0.11 & 13.1 & 10.6 & -78.4 \\
17:43:05.48 & -28:52:21.9 & 0.65 & 0.09 & 0.96 & 0.20 & 17.4 & 10.3 & 55.2 \\
17:43:45.60 & -28:58:56.5 & 0.77 & 0.10 & 1.49 & 0.30 & 17.1 & 15.7 & -13.0 \\
17:42:42.90 & -28:50:55.3 & 0.63 & 0.11 & 0.48 & 0.16 & 11.8 & 8.5 & 71.2 \\
17:44:02.60 & -29:02:14.0 & 1.99 & 0.17 & 3.11 & 0.39 & 17.4 & 11.4 & 63.8 \\
17:43:14.13 & -28:56:01.6 & 1.71 & 0.13 & 2.06 & 0.26 & 15.1 & 10.3 & 73.1 \\
17:43:54.72 & -29:01:32.4 & 1.11 & 0.10 & 1.50 & 0.22 & 15.5 & 11.6 & 62.3 \\
17:43:42.56 & -29:01:33.7 & 1.47 & 0.14 & 4.71 & 0.54 & 24.9 & 16.3 & 16.5 \\
17:43:42.79 & -29:02:26.6 & 0.96 & 0.11 & 3.99 & 0.57 & 25.5 & 22.5 & 71.0 \\
17:42:23.78 & -28:52:48.4 & 2.98 & 0.10 & 2.90 & 0.18 & 13.8 & 8.9 & 65.0 \\
17:42:13.82 & -28:52:10.8 & 0.64 & 0.08 & 0.86 & 0.16 & 16.4 & 10.3 & 64.7 \\
17:42:42.65 & -28:52:04.8 & 0.63 & 0.14 & 0.76 & 0.25 & 16.2 & 8.2 & 71.2 \\
17:43:17.04 & -29:02:12.3 & 1.13 & 0.13 & 1.29 & 0.25 & 15.0 & 9.5 & 76.0 \\
17:42:06.93 & -28:54:03.3 & 0.66 & 0.07 & 1.02 & 0.18 & 16.4 & 12.8 & 56.9 \\
17:42:53.59 & -29:00:53.9 & 0.82 & 0.08 & 1.00 & 0.17 & 15.0 & 10.5 & 63.4 \\
17:42:08.61 & -28:55:36.4 & 15.15 & 0.10 & 16.59 & 0.19 & 14.6 & 9.5 & 69.9 \\
17:42:03.50 & -28:55:05.0 & 1.02 & 0.07 & 1.16 & 0.14 & 14.8 & 9.7 & 62.5 \\
17:43:28.88 & -29:06:47.2 & 14.85 & 0.14 & 21.44 & 0.32 & 19.0 & 9.6 & 83.0 \\
17:42:31.07 & -28:59:45.9 & 7.80 & 0.11 & 10.44 & 0.24 & 15.8 & 11.0 & 64.7 \\
17:43:39.27 & -29:09:07.8 & 1.56 & 0.14 & 3.98 & 0.48 & 21.8 & 15.0 & 46.7 \\
17:42:41.76 & -29:02:10.2 & 2.09 & 0.13 & 2.10 & 0.22 & 14.0 & 9.1 & 69.1 \\
17:41:26.16 & -28:53:29.1 & 16.06 & 0.15 & 16.77 & 0.27 & 14.2 & 9.3 & 79.0 \\
17:42:57.26 & -29:07:30.7 & 0.67 & 0.12 & 0.78 & 0.24 & 15.0 & 9.9 & 56.7 \\
17:43:43.44 & -29:13:57.7 & 1.54 & 0.15 & 1.84 & 0.29 & 15.8 & 9.5 & 59.0 \\
17:43:42.83 & -29:14:33.8 & 1.11 & 0.23 & 1.88 & 0.55 & 18.8 & 10.9 & 61.9 \\
17:42:49.20 & -29:07:25.2 & 0.70 & 0.12 & 1.99 & 0.44 & 23.1 & 15.9 & 57.5 \\
17:41:42.85 & -28:59:02.8 & 2.90 & 0.09 & 3.02 & 0.15 & 14.4 & 9.0 & 71.9 \\
17:43:36.04 & -29:14:39.5 & 2.64 & 0.17 & 1.97 & 0.26 & 12.0 & 7.9 & 68.9 \\
17:42:30.64 & -29:05:59.7 & 1.67 & 0.10 & 4.47 & 0.35 & 22.1 & 15.9 & 60.6 \\
17:41:53.05 & -29:01:18.0 & 4.97 & 0.07 & 6.03 & 0.15 & 15.1 & 10.4 & 70.8 \\
17:43:34.86 & -29:15:04.1 & 0.78 & 0.14 & 0.75 & 0.25 & 13.4 & 9.3 & 77.3 \\
17:43:10.18 & -29:12:14.5 & 19.99 & 0.12 & 20.79 & 0.21 & 14.2 & 9.3 & 69.3 \\
17:44:03.76 & -29:19:53.3 & 2.90 & 0.34 & 4.71 & 0.77 & 18.6 & 10.2 & 70.7 \\
17:43:37.41 & -29:16:36.9 & 0.77 & 0.20 & 3.72 & 1.05 & 31.7 & 18.1 & 52.8 \\
17:44:09.04 & -29:21:33.3 & 4.23 & 0.61 & 3.58 & 0.97 & 12.8 & 8.4 & 73.1 \\
17:41:23.09 & -29:01:04.2 & 1.43 & 0.12 & 1.72 & 0.24 & 14.3 & 11.4 & 74.1 \\
17:42:38.84 & -29:12:18.0 & 2.42 & 0.11 & 2.98 & 0.22 & 14.6 & 11.3 & 80.8 \\
17:43:02.02 & -29:15:49.1 & 0.86 & 0.12 & 3.04 & 0.56 & 25.1 & 18.7 & 44.3 \\
17:43:53.40 & -29:23:30.2 & 1.79 & 0.32 & 3.25 & 0.85 & 17.7 & 13.7 & -87.7 \\
17:42:21.46 & -29:13:00.3 & 39.43 & 0.21 & 41.16 & 0.37 & 14.3 & 9.2 & 71.1 \\
17:41:59.99 & -29:10:47.5 & 4.28 & 0.15 & 4.17 & 0.26 & 13.8 & 8.9 & 67.7 \\
17:41:41.63 & -29:08:31.7 & 1.94 & 0.07 & 2.43 & 0.14 & 15.1 & 10.9 & 66.7 \\
17:42:11.83 & -29:13:29.7 & 18.93 & 0.13 & 19.05 & 0.22 & 14.0 & 9.1 & 69.3 \\
17:41:21.95 & -29:07:17.1 & 0.52 & 0.08 & 0.49 & 0.13 & 14.1 & 8.0 & 64.0 \\
17:42:31.10 & -29:16:42.3 & 19.80 & 0.14 & 23.02 & 0.26 & 15.1 & 9.7 & 69.3 \\
17:43:40.56 & -29:27:09.5 & 1.16 & 0.21 & 1.23 & 0.39 & 13.7 & 10.3 & 60.6 \\
17:41:30.70 & -29:11:03.0 & 5.72 & 0.10 & 6.05 & 0.18 & 14.3 & 9.4 & 66.1 \\
17:43:41.00 & -29:28:41.1 & 3.04 & 0.32 & 5.71 & 0.83 & 19.6 & 11.7 & 67.4 \\
17:43:02.31 & -29:23:42.3 & 1.13 & 0.08 & 1.62 & 0.19 & 16.2 & 11.6 & 65.3 \\
17:40:55.27 & -29:06:53.6 & 2.87 & 0.47 & 4.31 & 1.05 & 17.3 & 10.8 & 57.3 \\
17:41:24.27 & -29:10:52.6 & 1.42 & 0.17 & 6.59 & 0.96 & 27.7 & 22.7 & 76.1 \\
17:43:12.19 & -29:26:03.7 & 22.29 & 0.23 & 23.56 & 0.42 & 14.3 & 9.4 & 66.1 \\
17:42:53.79 & -29:24:57.0 & 0.90 & 0.12 & 2.51 & 0.43 & 22.5 & 16.3 & 55.8 \\
17:41:07.57 & -29:10:35.2 & 1.16 & 0.12 & 1.64 & 0.26 & 16.7 & 10.6 & 54.1 \\
17:41:45.42 & -29:16:20.8 & 4.05 & 0.08 & 5.40 & 0.16 & 16.1 & 10.5 & 63.1 \\
17:40:55.28 & -29:10:25.1 & 29.68 & 0.25 & 32.41 & 0.45 & 14.5 & 9.6 & 69.6 \\
17:42:10.67 & -29:20:54.0 & 1.39 & 0.13 & 1.48 & 0.23 & 14.3 & 9.6 & 58.7 \\
17:43:09.33 & -29:28:58.2 & 2.56 & 0.19 & 2.36 & 0.31 & 14.0 & 7.7 & 59.5 \\
17:43:10.20 & -29:29:52.0 & 3.29 & 0.38 & 3.94 & 0.75 & 15.2 & 10.0 & 62.7 \\
17:41:02.20 & -29:12:33.2 & 1.23 & 0.11 & 1.66 & 0.24 & 15.3 & 11.8 & 60.4 \\
17:41:28.40 & -29:16:34.3 & 0.69 & 0.08 & 1.32 & 0.23 & 19.1 & 12.9 & 79.4 \\
17:40:31.41 & -29:20:19.1 & 4.99 & 0.21 & 5.48 & 0.41 & 13.6 & 10.9 & 54.9 \\
\enddata
\tablecomments{Sources have greater than 5$\sigma$ peak brightness significance according to sfind and better than $\sim3\sigma$ confidence according to the AIPS SAD/JMFIT routines.  Columns 1--2 give the position of the source in J2000 coordinates; columns 3--4 give the peak flux and its uncertainty; columns 5--6 give the integrated flux and its uncertainty; columns 7--9 give the major and minor axes of the source and its position angle.}
\tablenotetext{a}{Position angle increases to the east from celestial north.}

\end{deluxetable}

\begin{deluxetable}{ccccccccc}
\tablecaption{Compact Sources in 20 cm Survey \label{ps20cm5}}
\tablewidth{0pt}
\tablehead{
\colhead{RA} & \colhead{Dec} & \colhead{$S_p$} & \colhead{$\sigma_{S_p}$} & \colhead{$S_i$} & \colhead{$\sigma_{S_i}$} & \colhead{$\theta_M$} & \colhead{$\theta_m$} & \colhead{$\theta_{PA}$ \tablenotemark{b}} \\  
\colhead{(J2000)} & \colhead{(J2000)} & \colhead{mJy beam$^{-1}$} & \colhead{mJy beam$^{-1}$} & \colhead{mJy} & \colhead{mJy} & \colhead{arcsec} & \colhead{arcsec} & \colhead{degrees} \\
}
\startdata
17:44:48.76 & -28:28:14.9 & 15.3 & 0.6 & 31.0 & 1.8 & 21.8 & 11.7 & 33.3 \\ 
17:44:45.93 & -28:35:14.3 & 8.2 & 0.6 & 6.9 & 1.0 & 13.3 & 8.0 & 56.5 \\ 
17:44:40.82 & -28:39:55.6 & 12.6 & 0.6 & 13.1 & 1.2 & 13.3 & 9.9 & 51.2 \\ 
17:44:23.01 & -28:38:51.8 & 5.6 & 0.6 & 14.6 & 2.1 & 21.8 & 15.2 & 28.2 \\ 
17:44:19.10 & -28:41:21.8 & 5.7 & 0.6 & 4.3 & 1.0 & 11.2 & 8.6 & 55.9 \\ 
17:43:45.72 & -28:32:06.0 & 5.4 & 0.6 & 6.2 & 1.2 & 13.8 & 10.4 & 93.3 \\ 
17:42:48.35 & -28:25:28.4 & 14.0 & 0.6 & 25.2 & 1.6 & 20.5 & 11.1 & 28.6 \\ 
17:42:39.71 & -28:41:40.8 & 4.5\tablenotemark{a} & 0.6 & 4.5 & 0.6 & 14.0 & 9.0 & 60 \\ 
17:41:28.60 & -28:36:55.5 & 5.2\tablenotemark{a} & 0.6 & 5.2 & 0.6 & 14.0 & 9.0 & 60 \\ 
17:41:19.06 & -28:24:19.7 & 9.2\tablenotemark{a} & 0.6 & 9.2 & 0.6 & 14.0 & 9.0 & 60 \\ 
17:41:26.20 & -28:53:30.0 & 32.9 & 0.6 & 47.7 & 1.4 & 16.3 & 11.2 & 53.4 \\ 
17:40:41.81 & -28:48:13.6 & 44.6 & 0.6 & 100.1 & 1.9 & 21.2 & 13.4 & 121.3 \\ 
17:40:35.33 & -28:41:42.4 & 5.3 & 0.6 & 7.7 & 1.4 & 18.9 & 9.6 & 131.8 \\ 
17:40:33.42 & -28:44:05.1 & 4.7 & 0.6 & 5.4 & 1.2 & 16.6 & 8.7 & 131.3 \\ 
17:40:07.99 & -28:42:04.5 & 55.8 & 0.6 & 246.3 & 3.1 & 39.5 & 14.1 & 126.9 \\ 
17:40:55.60 & -28:56:23.6 & 3.3 & 0.6 & 4.5 & 1.4 & 16.5 & 10.4 & 78.8 \\ 
17:40:58.39 & -28:57:31.9 & 7.9 & 0.6 & 10.6 & 1.3 & 17.0 & 9.9 & 70.5 \\ 
17:40:55.21 & -29:10:25.4 & 20.0 & 0.6 & 42.4 & 1.8 & 22.9 & 11.7 & 42.7 \\ 
17:40:20.49 & -29:13:44.4 & 8.1 & 0.6 & 11.3 & 1.4 & 14.6 & 12.0 & 56.4 \\ 
17:40:02.07 & -29:23:36.5 & 5.1 & 0.6 & 6.2 & 1.3 & 17.1 & 9.0 & 53.1 \\ 
17:39:39.58 & -29:31:22.6 & 4.1 & 0.6 & 11.6 & 2.2 & 23.9 & 14.8 & 47.9 \\ 
17:39:02.72 & -29:26:06.5 & 2.8 & 0.4 & 49.1 & 8.1 & 54.7 & 40.9 & 69.4 \\ 
17:40:31.36 & -29:20:19.1 & 5.2 & 0.6 & 6.0 & 1.2 & 15.3 & 9.5 & 61.3 \\ 
17:40:23.23 & -29:32:15.3 & 7.7 & 0.6 & 16.5 & 1.8 & 22.0 & 12.3 & 14.7 \\ 
17:40:54.43 & -29:29:46.8 & 41.1 & 0.6 & 102.1 & 2.0 & 19.7 & 15.9 & 157.1 \\ 
17:41:30.68 & -29:11:02.8 & 7.8 & 0.6 & 19.4 & 2.0 & 18.5 & 16.9 & 44.8 \\ 
17:43:14.19 & -28:56:01.3 & 3.3 & 0.6 & 3.3 & 1.1 & 13.4 & 9.5 & 70.7 \\ 
17:42:08.49 & -28:55:36.0 & 8.3 & 0.6 & 12.6 & 1.5 & 19.2 & 10.0 & 76.6 \\ 
17:42:31.21 & -28:59:45.2 & 4.3 & 0.6 & 7.1 & 1.5 & 19.0 & 11.0 & 77.1 \\ 
17:43:28.87 & -29:06:46.4 & 19.3 & 0.6 & 45.8 & 2.0 & 18.2 & 16.4 & 47.9 \\ 
17:43:10.11 & -29:12:13.2 & 6.1 & 0.6 & 9.6 & 1.5 & 18.7 & 10.6 & 41.7 \\
17:43:12.19 & -29:26:01.5 & 7.9 & 0.6 & 14.2 & 1.6 & 20.4 & 11.1 & 89.4 \\ 
17:44:05.86 & -29:17:42.8 & 8.4 & 0.6 & 50.1 & 4.0 & 29.2 & 25.8 & 0.0 \\ 
17:44:05.76 & -29:28:38.5 & 11.3 & 0.6 & 48.4 & 3.1 & 25.7 & 21.0 & 30.1 \\ 
17:44:57.17 & -28:38:02.7 & 6.4 & 0.6 & 18.0 & 2.2 & 20.8 & 17.1 & 94.4 \\ 
17:44:36.69 & -28:57:10.2 & 79.9 & 1.8 & 171.8 & 5.3 & 26.3 & 10.3 & 82.1 \\ 
17:41:42.87 & -28:59:02.1 & 4.1 & 0.6 & 2.9 & 0.9 & 11.5 & 7.5 & 67.0 \\ 
17:40:55.08 & -29:06:57.2 & 5.0 & 0.3 & 18.6 & 1.5 & 34.9 & 13.5 & 155.7 \\ 
17:41:45.52 & -29:16:20.2 & 3.4 & 0.6 & 8.0 & 2.0 & 21.1 & 14.2 & 92.4 \\ 
17:43:31.08 & -28:38:07.8 & 5.0 & 0.6 & 18.0 & 2.7 & 26.2 & 17.1 & 93.0 \\ 
17:41:27.92 & -28:52:51.7 & 8.2 & 0.6 & 17.1 & 1.8 & 18.3 & 14.4 & 52.0 \\ 
17:42:21.48 & -29:13:01.5 & 64.7 & 0.6 & 104.0 & 1.5 & 16.4 & 12.4 & 40.0 \\ 
17:42:11.87 & -29:13:31.0 & 19.1 & 0.6 & 28.5 & 1.4 & 14.5 & 12.9 & 35.0 \\ 
17:42:31.14 & -29:16:41.6 & 16.6 & 0.6 & 23.4 & 1.4 & 16.9 & 10.5 & 61.0 \\ 
17:42:00.06 & -29:10:49.4 & 5.1 & 0.6 & 9.3 & 1.6 & 17.3 & 13.3 & 159.0 \\ 
17:42:10.78 & -29:20:54.2 & 3.8 & 0.6 & 5.7 & 1.5 & 17.0 & 11.2 & 61.0 \\ 
17:43:03.32 & -28:50:57.5 & 6.6 & 0.6 & 13.7 & 1.8 & 17.2 & 15.1 & 144.0 \\ 
17:43:02.80 & -28:51:12.6 & 5.7 & 0.6 & 8.6 & 1.5 & 14.7 & 13.0 & 2.0 \\ 
\enddata
\tablecomments{Sources have greater than 5$\sigma$ peak brightness significance according to the AIPS SAD/JMFIT routines.  Columns 1--2 give the position of the source in J2000 coordinates; columns 3--4 give the peak flux and its uncertainty; columns 5--6 give the integrated flux and its uncertainty; columns 7--9 give the major and minor axes of the source and its position angle.}
\tablenotetext{a}{Source parameters found by fixing the source size to the beam size of 14\arcsec\ by 9\arcsec.  This assumption may result in the peak brightness being overestimated by about 50\%.}
\tablenotetext{b}{Position angle increases to the east from celestial north.}

\end{deluxetable}

\begin{deluxetable}{cccccccc}
\tablecaption{Compact Sources with 6 cm Polarization \label{6poln}}
\tablewidth{0pt}
\tablehead{
\colhead{RA$_6$} & \colhead{Dec$_6$} & \colhead{$S_i$} & \colhead{$\sigma_{S_i}$} & \colhead{$pS_i$} & \colhead{$\sigma_{pS_i}$} & \colhead{$f_{pol}$} & \colhead{$\sigma_{f_{pol}}$} \\  
\colhead{(J2000)} & \colhead{(J2000)} & \colhead{mJy} & \colhead{mJy} & \colhead{mJy} & \colhead{mJy} & \colhead{} & \colhead{} \\  
}
\startdata
17:40:55.28  & -29:10:25.1  & 32.4 & 0.5 & 3.4 & 0.3 &  0.10 &  0.01 \\  
17:42:21.46  & -29:13:00.3  & 41.2 & 0.4 & 1.5 & 0.3 &  0.04 &  0.01 \\  
\enddata
\tablecomments{Columns 1--2 give the position of the 6 cm source in J2000 coordinates; columns 3--4 give the integrated flux and its uncertainty; columns 5--6 give the integrated polarized flux and its uncertainty; columns 7--8 give the polarization fraction and its error.}

\end{deluxetable}

\begin{deluxetable}{cccccccc}
\tablecaption{Point Sources with 20 cm Polarization \label{20poln}}
\tablewidth{0pt}
\tablehead{
\colhead{RA$_{20}$} & \colhead{Dec$_{20}$} & \colhead{$S_i$} & \colhead{$\sigma_{S_i}$} & \colhead{$pS_i$} & \colhead{$\sigma_{pS_i}$} & \colhead{$f_{pol}$} & \colhead{$\sigma_{f_{pol}}$} \\  
\colhead{(J2000)} & \colhead{(J2000)} & \colhead{mJy beam$^{-1}$} & \colhead{mJy beam$^{-1}$} & \colhead{mJy} & \colhead{mJy} & \colhead{} & \colhead{} \\  
}
\startdata
17:41:26.20  &  -28:53:30.0  &  47.7  & 1.4 &  9.1 & 2.2 & 0.190 & 0.045 \\ 
17:40:41.81  &  -28:48:13.6  &  100.1  & 1.9 &  3.8\tablenotemark{a} & 0.28 & 0.038 & 0.003 \\ 
17:40:07.99   &  -28:42:4.5   &  246.3  & 3.1 &  6.1\tablenotemark{a} & 0.28 & 0.025 & 0.001 \\ 
17:40:54.43  &  -29:29:46.8  &  102.1  & 2.0 & 24.8 & 2.8 & 0.243 & 0.028 \\ 
17:42:21.48  &  -29:13:1.5   &  104.0  & 1.5 & 22.7 & 2.8 & 0.218 & 0.027 \\ 
17:42:11.87  &  -29:13:31.0  &  28.5  & 1.4 &  1.6\tablenotemark{a} & 0.28 & 0.056 & 0.010 \\ 
17:42:31.14  &  -29:16:41.6  &  23.4  & 1.4 &  1.2\tablenotemark{a} & 0.28 & 0.051 & 0.012 \\ 
\enddata 
\tablenotetext{a}{Source parameters found by fixing the source size to the beam size of that field.}
\tablecomments{Columns 1--2 give the position of the 20 cm source in J2000 coordinates; columns 3--4 give the integrated 20 cm flux and its uncertainty; columns 5--6 give the integrated 20 cm polarized flux and its uncertainty; columns 7--8 give the polarization fraction and its error.}

\end{deluxetable}

\begin{deluxetable}{cccccc}
\tablecaption{Compact Sources in both 6 and 20 cm Catalogs \label{ps6-20}}
\tablewidth{0pt}
\tablehead{
\colhead{RA$_6$} & \colhead{Dec$_6$} & \colhead{RA$_{20}$} & \colhead{Dec$_{20}$} & \colhead{$\alpha$ \tablenotemark{a}} & \colhead{$\sigma_\alpha$} \\
\colhead{(J2000)} & \colhead{(J2000)} & \colhead{(J2000)} & \colhead{(J2000)} & \colhead{} & \colhead{} \\
}
\startdata
17:42:07.92 & -29:18:51.9 & 17:42:07.83 & -29:18:48.0 & -1.1 & 0.4 \\
17:40:31.41 & -29:20:19.1 & 17:40:31.36 & -29:20:19.1 & -0.1 & 0.2 \\
17:40:55.27 & -29:06:53.6 & 17:40:55.08 & -29:06:57.2 & -1.2 & 0.2 \\
17:40:55.28 & -29:10:25.1 & 17:40:55.21 & -29:10:25.4 & -0.22 & 0.04 \\
17:41:30.70 & -29:11:03.0 & 17:41:30.68 & -29:11:02.8 & -1.0 & 0.1 \\
17:41:45.42 & -29:16:20.8 & 17:41:45.52 & -29:16:20.2 & -0.3 & 0.2 \\
17:41:59.99 & -29:10:47.5 & 17:42:00.06 & -29:10:49.4 & -0.7 & 0.2 \\
17:42:10.67 & -29:20:54.0 & 17:42:10.78 & -29:20:54.2 & -1.1 & 0.2 \\
17:42:31.10 & -29:16:42.3 & 17:42:31.14 & -29:16:41.6 & -0.0 & 0.1 \\
17:42:11.83 & -29:13:29.7 & 17:42:11.87 & -29:13:31.0 & -0.33 & 0.04 \\
17:42:21.46 & -29:13:00.3 & 17:42:21.48 & -29:13:01.5 & -0.76 & 0.01 \\
17:43:12.19 & -29:26:03.7 & 17:43:12.19 & -29:26:01.5 & 0.4 & 0.1 \\
17:43:10.18 & -29:12:14.5 & 17:43:10.11 & -29:12:13.2 & 0.6 & 0.1 \\
17:43:28.88 & -29:06:47.2 & 17:43:28.87 & -29:06:46.4 & -0.61 & 0.04 \\
17:44:37.05 & -28:57:09.5 & 17:44:36.69 & -28:57:10.2 & -0.94 & 0.04 \\
17:43:14.13 & -28:56:01.6 & 17:43:14.19 & -28:56:01.3 & -0.4 & 0.3 \\
17:43:02.75 & -28:51:09.2 & 17:43:02.80 & -28:51:12.6 & -0.5 & 0.2 \\
17:43:03.46 & -28:50:55.4 & 17:43:03.32 & -28:50:57.5 & -1.5 & 0.1 \\
17:42:39.65 & -28:41:44.5 & 17:42:39.71 & -28:41:40.8 & -0.7 & 0.1 \\
17:41:53.05 & -29:01:18.0 & 17:41:52.99 & -29:01:16.6 & -0.0 & 0.2 \\
17:41:42.85 & -28:59:02.8 & 17:41:42.87 & -28:59:02.1 & 0.0 & 0.3 \\
17:42:08.61 & -28:55:36.4 & 17:42:08.49 & -28:55:36.0 & 0.2 & 0.1 \\
17:42:31.07 & -28:59:45.9 & 17:42:31.21 & -28:59:45.2 & 0.3 & 0.2 \\
17:41:26.16 & -28:53:29.1 & 17:41:26.20 & -28:53:30.0 & -0.85 & 0.03 \\
17:43:31.22 & -28:38:07.7 & 17:43:31.08 & -28:38:07.8 & -0.9 & 0.1 \\
17:43:20.50 & -28:36:33.4 & 17:43:20.64 & -28:36:32.2 & -1.3 & 0.3 \\
17:43:45.58 & -28:32:05.9 & 17:43:45.72 & -28:32:06.0 & 0.5 & 0.2 \\
17:44:18.97 & -28:41:24.3 & 17:44:19.10 & -28:41:21.8 & 0.4 & 0.2 \\
17:44:22.88 & -28:38:51.2 & 17:44:23.01 & -28:38:51.8 & -0.1 & 0.1 \\
17:44:40.87 & -28:39:56.2 & 17:44:40.82 & -28:39:55.6 & 0.3 & 0.1 \\
17:44:45.92 & -28:35:15.2 & 17:44:45.93 & -28:35:14.3 & 0.5 & 0.1 \\
17:44:57.38 & -28:38:02.3 & 17:44:57.17 & -28:38:02.7 & -0.3 & 0.1 \\
17:44:48.78 & -28:28:13.2 & 17:44:48.76 & -28:28:14.9 & -0.9 & 0.1 \\
\enddata
\tablecomments{Comparing all sources with greater than $3\sigma$ confidence according to the AIPS SAD/JMFIT routines.  Columns 1--2 give the position of the 6 cm source in J2000 coordinates; columns 3--4 give the position of the 20 cm source; columns 5--6 give the 6/20 cm spectral index from the integrated fluxes of the source and its uncertainty.}
\tablenotetext{a}{Assumes  $S_\nu\propto\nu^{\alpha}$.}

\end{deluxetable}

\begin{deluxetable}{ccc}
\tablecaption{Compact Sources in 6 cm Catalog Only \label{ps6no20}}
\tablewidth{0pt}
\tablehead{
\colhead{RA$_6$} & \colhead{Dec$_6$} & \colhead{min $\alpha$ \tablenotemark{a}} \\
\colhead{(J2000)} & \colhead{(J2000)} & \colhead{} \\
}
\startdata
17:43:34.26 & -28:30:42.6 & -0.6 \\
17:43:41.63 & -28:32:47.9 & -0.7 \\
17:43:42.47 & -28:33:28.1 & -0.2 \\
17:44:54.55 & -28:44:59.1 & -0.3 \\
17:44:31.06 & -28:47:16.4 & 0.1 \\
17:43:47.70 & -28:41:49.4 & 0.8 \\
17:43:39.77 & -28:41:51.7 & -0.7 \\
17:43:17.04 & -28:43:01.7 & 0.2 \\
17:43:22.96 & -28:44:52.3 & -0.9 \\
17:43:12.35 & -28:44:33.7 & -0.9 \\
17:43:10.12 & -28:45:17.3 & -1.4 \\
17:43:23.12 & -28:47:54.0 & -0.5 \\
17:42:28.74 & -28:41:48.9 & -0.2 \\
17:42:44.73 & -28:45:48.7 & 0.2 \\
17:42:44.52 & -28:45:37.7 & -0.0 \\
17:42:04.91 & -28:42:10.4 & 0.4 \\
17:43:46.90 & -28:57:05.4 & -0.6 \\
17:42:24.48 & -28:45:57.6 & -1.0 \\
17:43:05.48 & -28:52:21.9 & -0.3 \\
17:43:45.60 & -28:58:56.5 & -0.2 \\
17:42:42.90 & -28:50:55.3 & -0.9 \\
17:44:02.60 & -29:02:14.0 & 0.3 \\
17:43:54.72 & -29:01:32.4 & -0.6 \\
17:43:42.56 & -29:01:33.7 & 0.7 \\
17:43:42.79 & -29:02:26.6 & 0.3 \\
17:42:23.78 & -28:52:48.4 & 0.6 \\
17:42:13.82 & -28:52:10.8 & -0.1 \\
17:42:42.65 & -28:52:04.8 & -0.2 \\
17:43:17.04 & -29:02:12.3 & -0.2 \\
17:42:06.93 & -28:54:03.3 & 0.3 \\
17:42:53.59 & -29:00:53.9 & -0.1 \\
17:42:03.50 & -28:55:05.0 & 0.1 \\
17:43:39.27 & -29:09:07.8 & 0.3 \\
17:42:41.76 & -29:02:10.2 & 0.4 \\
17:42:57.26 & -29:07:30.7 & -0.5 \\
17:43:43.44 & -29:13:57.7 & -0.5 \\
17:43:42.83 & -29:14:33.8 & -0.5 \\
17:42:49.20 & -29:07:25.2 & 0.3 \\
17:43:36.04 & -29:14:39.5 & -0.4 \\
17:42:30.64 & -29:05:59.7 & 1.0 \\
17:43:34.86 & -29:15:04.1 & -1.1 \\
17:44:03.76 & -29:19:53.3 & -0.1 \\
17:43:37.41 & -29:16:36.9 & 0.2 \\
17:44:09.04 & -29:21:33.3 & -0.2 \\
17:41:23.09 & -29:01:04.2 & 0.4 \\
17:42:38.84 & -29:12:18.0 & 0.9 \\
17:43:02.02 & -29:15:49.1 & 0.4 \\
17:43:53.40 & -29:23:30.2 & 1.3 \\
17:41:41.63 & -29:08:31.7 & 0.8 \\
17:41:21.95 & -29:07:17.1 & -0.7 \\
17:43:40.56 & -29:27:09.5 & -0.8 \\
17:43:41.00 & -29:28:41.1 & 0.1 \\
17:43:02.31 & -29:23:42.3 & -0.2 \\
17:41:24.27 & -29:10:52.6 & 1.6 \\
17:42:53.79 & -29:24:57.0 & 0.5 \\
17:41:07.57 & -29:10:35.2 & 0.6 \\
17:43:09.33 & -29:28:58.2 & -0.4 \\
17:43:10.20 & -29:29:52.0 & -0.2 \\
17:41:02.20 & -29:12:33.2 & 0.7 \\
17:41:28.40 & -29:16:34.3 & 0.5 \\
\enddata
\tablenotetext{a}{Assumes  $S_\nu\propto\nu^{\alpha}$.}
\tablecomments{Columns 1--2 give the position of the 6 cm source in J2000 coordinates; column 3 gives the limit on the 6/20 cm spectral index.}

\end{deluxetable}

\begin{deluxetable}{ccccccc}
\tablecaption{20 cm Sources Matching the Catalog of YHC04 with $b>$0\ddeg1 \label{yhc}}
\tablewidth{0pt}
\tablehead{
\colhead{RA$_{20}$} & \colhead{Dec$_{20}$} & \colhead{$S_i$} & \colhead{$\sigma_{S_i}$} & \colhead{Previous Name} & \colhead{$S$} & \colhead{$\sigma_{S}$} \\  
\colhead{(J2000)} & \colhead{(J2000)} & \colhead{mJy} & \colhead{mJy} & \colhead{} & \colhead{mJy} & \colhead{mJy} \\
}
\startdata
17:44:40.82 & -28:39:55.6 & 13.1 & 1.2 & G0.12+0.32   & 17.0 & 2.8 \\
17:44:19.10 & -28:41:21.8 & 4.3 & 0.9 & G0.06+0.37   & 4.8 & 1.5 \\
17:44:36.69 & -28:57:10.2 & 171.8 & 5.3 & G359.87+0.18 & 153.7  & 1.9 \\
17:43:12.19 & -29:26:01.5 & 14.2 & 1.6 & G359.29+0.19 & 17.8 & 1.8 \\
17:43:28.87 & -29:06:46.4 & 45.8 & 2.0 & G359.60+0.31 & 55.2 & 2.9 \\
17:43:10.11 & -29:12:13.2 & 17.1 & 2.4 & G359.49+0.32 & 9.6  & 1.7 \\
17:42:31.14 & -29:16:41.6 & 23.4 & 1.4 & G359.35+0.40 & 22.5  & 2.0 \\
17:42:10.78 & -29:20:54.2 & 5.7 & 1.5 & G359.26+0.42 & 5.0 & 2.0 \\
17:42:21.48 & -29:13:01.5 & 104.0 & 1.5 & G359.39+0.46 & 102.0 & 1.8 \\
17:42:11.87 & -29:13:31.0 & 28.5 & 1.4 & G359.36+0.49 & 32.2 & 1.9 \\
17:41:45.52 & -29:16:20.2 & 8.0 & 2.0 & G359.27+0.54 & 8.5  & 2.1 \\
17:40:54.43 & -29:29:46.8 & 102.1 & 2.0 & G358.98+0.58 & 103.3 & 2.4 \\
17:41:30.68 & -29:11:02.8 & 19.4 & 2.0 & G359.32+0.63 & 20.1 & 2.0 \\
17:40:55.21 & -29:10:25.4 & 42.4 & 1.8 & G359.26+0.75 & 33.8 & 2.2 \\
17:42:00.06 & -29:10:49.4 & 9.3 & 1.6 & G359.38+0.55 & 14.1 & 2.0 \\
\enddata
\tablecomments{Columns 1--2 give the position of the 20 cm source in J2000 coordinates; columns 3--4 give the integrated 20 cm flux and its uncertainty; column 5 gives the name of the source in the survey of YHC04;  columns 6--7 give the integrated 20 cm flux measured by YHC04 and its uncertainty.}

\end{deluxetable}

\begin{deluxetable}{ccccccc}
\tablecaption{20 cm Sources Matching the Catalog of \citet{l98} \label{1lc}}
\tablewidth{0pt}
\tablehead{
\colhead{RA$_{20}$} & \colhead{Dec$_{20}$} & \colhead{$S_i$} & \colhead{$\sigma_{S_i}$} & \colhead{Previous Name} & \colhead{$S_{i1658}$} & \colhead{$S_{i1281}$} \\  
\colhead{(J2000)} & \colhead{(J2000)} & \colhead{mJy} & \colhead{mJy} & \colhead{} & \colhead{mJy} & \colhead{mJy} \\
}
\startdata
17:41:16.05 & -29:26:06.7 & 3.1 & 0.8 & 359.076+0.547 & 2.7 & 2.9 \\
17:44:48.76 & -28:28:14.9 & 31.1 & 1.8 & 0.305+0.394 & 9.4 & 12.8 \\
17:40:55.21 & -29:10:25.4 & 42.4 & 1.8 & 359.259+0.749 & 36.5 & 39.4 \\
17:40:31.36 & -29:20:19.1 & 6.0 & 1.2 & 359.073+0.735 & 5.3 & 4.3 \\
17:44:36.69 & -28:57:10.2 & 171.8 & 5.3 & 359.872+0.178 & 83.1 & 81.2 \\
17:42:21.48 & -29:13:01.5 & 104.0 & 1.5 & 359.388+0.460 & 60.6 & 155.8 \\
17:41:46.04 & -29:29:16.3 & 9.0 & 2.3 & 359.088+0.426 & 5.0 & 3.6 \\
\enddata
\tablecomments{Columns 1--2 give the position of the 20 cm source in J2000 coordinates; columns 3--4 give the integrated 20 cm flux and its uncertainty; column 5 gives the name of the source in the survey of \citet{l98};  columns 6--7 give the integrated fluxes measured in two bands near 20 cm by \citet{l98}.}

\end{deluxetable}

\begin{deluxetable}{cccccc}
\tablecaption{6 cm Sources Matching the Catalog of \citet{b94} \label{gpsr5}}
\tablewidth{0pt}
\tablehead{
\colhead{RA$_6$} & \colhead{Dec$_6$} & \colhead{$S_{i6}$} & \colhead{$\sigma_{S_{i6}}$} & \colhead{Previous Name} & \colhead{$S_{i}$} \\  
\colhead{(J2000)} & \colhead{(J2000)} & \colhead{mJy} & \colhead{mJy} & \colhead{} & \colhead{mJy} \\  
}
\startdata
17:44:48.78 & -28:28:13.2 & 10.9 & 0.7 & 0.306+0.394 & 6.8 \\
17:44:45.91 & -28:35:15.2 & 12.7 & 0.3 & 0.201+0.342 & 5.2 \\
17:44:57.38 & -28:38:02.3 & 12.7 & 0.7 & 0.185+0.282 & 4.5 \\
17:44:40.87 & -28:39:56.2 & 19.2 & 0.5 & 0.127+0.317 & 13.6 \\
17:44:18.97 & -28:41:24.3 & 6.8 & 0.2 & 0.059+0.372 & 5.9 \\
17:44:37.05 & -28:57:09.5 & 54.1 & 2.4 & 359.873+0.178 & 37.2 \\
17:43:28.88\tablenotemark{a} & -29:06:47.5 & 21.4 & 0.3 & 359.606+0.305 & 7.6 \\
17:43:28.88\tablenotemark{a} & -29:06:47.5 & 21.4 & 0.3 & 359.604+0.307 & 3.3 \\
17:43:10.18 & -29:12:14.5 & 20.8 & 0.2 & 359.491+0.316 & 10.1 \\
17:42:31.10 & -29:16:42.3 & 23.0 & 0.3 & 359.354+0.398 & 17.4 \\
17:43:12.19 & -29:26:03.7 & 23.6 & 0.4 & 359.299+0.189 & 17.9 \\
17:43:09.33 & -29:28:58.2 & 2.4 & 0.3 & 359.253+0.172 & 1.4 \\
17:43:10.20 & -29:29:52.0 & 4.0 & 0.8 & 359.242+0.162 & 1.5 \\
\enddata
\tablenotetext{a}{Two sources in \citet{b94} are unresolved here.}
\tablecomments{Columns 1--2 give the position of the 6 cm source in J2000 coordinates; columns 3--4 give the integrated 6 cm flux and its uncertainty; column 5 gives the name of the source in the survey of \citet{b94};  column 6 gives the integrated 6 cm flux measured by \citet{b94}.}

\end{deluxetable}

\begin{deluxetable}{cccccc}
\tablecaption{Spectral Indices for Sources in the Catalog of \citet{n04} \label{gcps}}
\tablewidth{0pt}
\tablehead{
\colhead{RA$_{20}$} & \colhead{Dec$_{20}$} & \colhead{$\alpha_{20/90}$} & \colhead{$\sigma_{\alpha_{20/90}}$} & \colhead{$\alpha_{6/20}$} & \colhead{$\sigma_{\alpha_{6/20}}$} \\  
\colhead{(J2000)} & \colhead{(J2000)} & \colhead{} & \colhead{} & \colhead{} & \colhead{} \\  
}
\startdata
17:44:48.755 & -28:28:14.85 & 0.33 & 0.06 & 0.85 & 0.07 \\
17:42:48.347 & -28:25:28.35 & 0.77 & 0.07 & -- & -- \\
17:41:49.359 & -28:43:20.45 & 1.22 & 0.36 & -- & -- \\
17:41:26.196 & -28:53:30.04 & 0.31 & 0.08 & 0.85 & 0.03 \\
17:40:41.808 & -28:48:13.61 & 0.62 & 0.03 & -- & -- \\
17:40:33.419 & -28:44:05.08 & 1.53 & 0.20 & -- & -- \\
17:40:07.992 & -28:42:04.50 & 0.12 & 0.03 & -- & -- \\
17:40:20.487 & -29:13:44.37 & 0.66 & 0.15 & -- & -- \\
17:40:54.433 & -29:29:46.78 & 0.32 & 0.02 & -- & -- \\
17:43:28.869 & -29:06:46.42 & 0.83 & 0.04 & 0.61 & 0.04 \\
17:44:36.688 & -28:57:10.19 & 0.74 & 0.03 & 0.94 & 0.04 \\
17:42:21.477 & -29:13:01.49 & -0.20 & 0.12 & 0.76 & 0.01 \\
17:43:03.318 & -28:50:57.45 & 0.08 & 0.25 & 1.50 & 0.14 \\
17:43:02.797 & -28:51:12.56 & 0.40 & 0.26 & 0.54 & 0.15 \\
\enddata
\tablecomments{Columns 1--2 give the position of the 20 cm source in J2000 coordinates; columns 3--4 give the 20/90 cm spectral index and its uncertainty; columns 5--6 give the 6/20 cm spectral index and its uncertainty.}

\end{deluxetable}

\clearpage

\clearpage

\clearpage

\clearpage

\clearpage

\clearpage

\begin{thebibliography}{}
\bibitem[Anantharamaiah et al.(1991)]{a91} Anantharamaiah, K. R. et al. 1991, MNRAS 249, 262

\bibitem[Bajaja \& van Albada(1979)]{b79} Bajaja, E. \& van Albada, G. D. 1979, A\&A, 75, 251

\bibitem[Bally \& Yusef-Zadeh(1989)]{b89} Bally, J. \& Yusef-Zadeh, F. 1989, ApJ, 336, 173

\bibitem[Beck(2001)]{b01} Beck, R. 2001, SSRv, 99, 243

\bibitem[Becker et al.(1994)]{b94} Becker, R. H., White, R. L., Helfand, D. J., \& Zoonematkermani, S. 1994 ApJS, 91, 347

\bibitem[Benford(1988)]{b88} Benford, G. 1988, ApJ, 333, 735

\bibitem[Bland-Hawthorn \& Cohen(2003)]{b03} Bland-Hawthorn, J \& Cohen, M. 2003, ApJ, 582, 246

\bibitem[Boldyrev \& Yusef-Zadeh(2006)]{b06} Boldyrev, S. \& Yusef-Zadeh, F. 2006, ApJ, 637L, 101

\bibitem[Blundell \& Rawlings(2000)]{b00} Blundell, K. M. \& Rawlings, S. 2000, AJ, 119, 1111

\bibitem[Bridle \& Schwab(1999)]{b99} Bridle, A. H. \& Schwab, F. S. 1999, Synthesis Imaging in Radio Astronomy II, ASP, 180, 371

\bibitem[Brentjens \& de Bruyn(2005)]{b05} Brentjens, M. A. \& de Bruyn, A. G. 2005, A\&A, 441, 1217

\bibitem[Brogan et al.(2003)]{br03} Brogan, C. L., Nord, M., Kassim, N., Lazio, J., \& Anantharamaiah, K. 2003, ANS, 324, 17

\bibitem[Burn(1966)]{b66} Burn, B. J., 1966, MNRAS, 133, 67

\bibitem[Condon et al.(1998)]{c98} Condon, J. J. et al.\ 1998, AJ, 115, 1693

\bibitem[Cordes \& Lazio(2004)]{c04} Cordes, J. M. \& Lazio, T. J. W. 2004, astro-ph/0207156

\bibitem[Cordes \& Rickett(1998)]{co98} Cordes, J. M. \& Rickett, B. J. 1998, ApJ, 507, 846

\bibitem[Cotton(1994)]{c94} Cotton, W. D. 1994, AIPS memo 86, \url{http://www.aips.nrao.edu/aipsmemo.html}

\bibitem[Cotton(1999)]{c99} Cotton, W. D. 1999, Synthesis Imaging in Radio Astronomy II, ASP, 180, 120

\bibitem[Davies et al.(1976)]{d76} Davies, R. D., Walsh, D., \& Booth, R. S. 1976, MNRAS, 177, 319

\bibitem[Downes et al.(1979)]{d79} Downes, D., Goss, W. M., Schwarz, U. J., \& Wouterloot, J. G. A. 1979, A\&AS, 35, 1

\bibitem[Downes \& Martin(1971)]{d71} Downes, D. \& Martin, A. H. M. 1971, Nature, 233, 112

\bibitem[Ekers et al(1975)]{e75} Ekers, R. D., Goss, W. M., Schwarz, U. J., Downes, D., \& Rogstad, D. H. 1975, A\&A, 43, 159

\bibitem[Fanaroff \& Riley(1974)]{f74} Fanaroff, B. L. \& Riley, J. M. 1974, MNRAS, 167, 31

\bibitem[Gelfand et al.(2004)]{g04} Gelfand, J. D., Lazio, T. J. W., \& Gaensler, B. M. 2004, ApJS, 155, 89

\bibitem[Gelfand et al.(2005)]{ge05} Gelfand, J. D., Lazio, T. J. W., \& Gaensler, B. M. 2005, ApJS, 159, 242

\bibitem[Giveon et al.(2005a)]{g05a} Giveon, U., Becker, R. H.,  Helfand, D. J., \& White, R. L. 2005, AJ, 129, 348

\bibitem[Giveon et al.(2005b)]{g05b} Giveon, U., Becker, R. H.,  Helfand, D. J., \& White, R. L. 2005, AJ, 130, 156


\bibitem[Gray(1994)]{g94} Gray, A. D. 1994, MNRAS, 270, 835

\bibitem[Gray et al.(1993)]{g93} Gray, A. D., et al. 1993, MNRAS, 264, 678

\bibitem[Haverkorn et al.(2000)]{h00} Haverkorn, M., Katgert, P., \& de Bruyn, A. G. 2000, A\&A, 356, L13

\bibitem[Haverkorn et al.(2004)]{h04} Haverkorn, M., Katgert, P., \& de Bruyn, A. G. 2004, A\&A, 427, 549

\bibitem[Heyvaerts et al.(1988)]{h88} Heyvaerts, J., Norman, C., \& Pudritz, R. E. 1988, ApJ, 330, 718

\bibitem[Helfand \& Becker(1987)]{h87} Helfand, D. J. \& Becker, R. H. 1987, ApJ, 314, 203

\bibitem[Hughes et al.(2007)]{h07} Hughes, A., Staveley-Smith, L., Kim, S., Wolleben, M., \& Filipovi\'c, M. 2007, accepted to MNRAS, astro-ph/0709.1990

\bibitem[Isaacman(1981)]{i81} Isaacman, R. 1981, A\&AS, 43, 405

\bibitem[Jaffe \& Perola(1973)]{j73} Jaffe, W. J. \& Perola, G. C. 1973, A\&A, 26, 423

\bibitem[Kaplan et al.(2000)]{k00} Kaplan, D. L., Cordes, J. M., \& Condon, J. J. 2000, ApJS, 126, 37

\bibitem[Lang et al.(1999a)]{l99a} Lang, C. C., Anantharamaiah, K. R., Kassim, N. E., \& Lazio, T. J. W. 1999, ApJ, 521, L41

\bibitem[Lang \& Anantharamaiah(2000)]{la00} Lang, C. C. \& Anantharamaiah, K. R. 2000, IAUJD, 14, 25

\bibitem[Lang et al.(1999b)]{l99} Lang, C. C., Morris, M., Echevarria, L. 1999, ApJ, 526, 727

\bibitem[Lang(1971)]{l71} Lang, K. R. 1971, ApJ, 164, 249

\bibitem[LaRosa et al.(2005)]{la05} LaRosa, T. N., Brogan, C. L., Shore, S. N., Lazio, T. J., Kassim, N. E., \& Nord, M. E.  2005, ApJ, 626, L23

\bibitem[LaRosa et al.(2000)]{l00} LaRosa, T. N., Kassim, N. E., Lazio, T. J. W., \& Hyman, S. D. 2000, AJ, 119, 207

\bibitem[LaRosa et al.(2001)]{l01} LaRosa, T. N., Lazio, T. J. W., \& Kassim, N. E.2001, ApJ, 563, 163

\bibitem[LaRosa et al.(2004)]{l04} LaRosa, T. N., Nord, M. E., J., T., Lazio, W., \& Kassim, N. E.  2004, ApJ, 607, 302

\bibitem[Law(2007)]{thesis} Law, C. J. 2007, PhD thesis, astro-ph/0705.0114

\bibitem[Law et al.(2008a)]{gcsurvey_gbt} Law, C. J., Yusef-Zadeh, F., Cotton, W. D., \& Maddalena, R. 2008, ApJS, in press, astro-ph/0801.4294


\bibitem[Law et al.(2008b)]{gcl_all} Law, C. J., Yusef-Zadeh, F., et al. 2008, in preparation

\bibitem[Lazio \& Cordes(1998a)]{l98} Lazio, T. J. W. \& Cordes, J. M. 1998, ApJS, 118, 201

\bibitem[Lazio \& Cordes(1998b)]{l98b} Lazio, T. J. W. \& Cordes, J. M. 1998, ApJ, 505, 715

\bibitem[Liszt \& Spiker(1995)]{l95} Liszt, H. S. \& Spiker, R. W. 1995, ApJS, 98, 259

\bibitem[Lorimer et al.(1995)]{lo95} Lorimer, D. R., Yates, J. A., Lyne, A. G., \& Gould, D. M. 1995, MNRAS, 273, 411

\bibitem[Lu et al.(2003)]{l03} Lu, F. J., Wang, Q. D., \& Lang, C. C. 2003, AJ, 126, 319

\bibitem[Manchester et al.(2005)]{ma05} Manchester, R. N., Hobbs, G. B., Teoh, A. \& Hobbs, M. 2005, AJ, 129, 1993 (See also \url{http://www.atnf.csiro.au/research/pulsar/psrcat})


\bibitem[Morris \& Serabyn(1996)]{m96} Morris, M. \& Serabyn, E. 1996, ARA\&A, 34, 645

\bibitem[Morris et al.(2006)]{m06} Morris, M., Uchida, K., \& Do, T. 2006, Nature, 440, 308

\bibitem[Muno et al.(2006)]{mu06} Muno, M. P., Bauer, F. E., Bandyopadhyay, R. M., \& Wang, Q. D. 2006, ApJS, 165, 173

\bibitem[Muno et al.(2005)]{mu05} Muno, M. P., Pfahl, E., Baganoff, F. K., Brandt, W. N., Ghez, A., Lu, J., \& Morris, M. R. 2005, ApJ, 622, L113

\bibitem[Nord et al.(2004)]{n04} Nord, M. E., Lazio, T. J. W., Kassim, N. E., Hyman, S. D., LaRosa, T. N., Brogan, C. L., \& Duric, N. 2004, AJ, 128, 1646 (erratum 131, 1886, [2006])

\bibitem[O'Dea(1998)]{o98} O'Dea, C. P. 1998, PASP, 110, 493

\bibitem[Pohl \& Schlickeiser(1992)]{p92} Pohl, M. \& Schlickeiser, R. 1992, A\&A, 262, 441


\bibitem[Reid(1993)]{r93} Reid, M. J. 1993, ARA\&A, 31, 345

\bibitem[Roy(2003)]{r03} Roy, S. 2003, A\&A, 403, 917

\bibitem[Roy et al.(2005)]{r05} Roy, S., Rao, A. P., \& Subrahmanyan, R. 2005, MNRAS, 360, 1305

\bibitem[Rybicki \& Lightman(1979)]{r79} Rybicki, G. B. \& Lightman, A. P. 1979, Radiative Processes in Astrophysics, John Wiley \& Sons, Inc., New York

\bibitem[Serabyn \& Morris(1994)]{s94} Serabyn, E. \& Morris, M. 1994, ApJ, 424, L91

\bibitem[Shore \& LaRosa(1999)]{s99} Shore, S. N. \&  LaRosa, T. N. 1999, ApJ, 521, 587

\bibitem[Sofue \& Handa(1984)]{s84} Sofue, Y. \& Handa, T. 1984, Nature, 310, 568

\bibitem[Sofue et al.(1992)]{so92} Sofue, Y., Murata, Y., \& Reich, W. 1992, PASJ, 44, 367

\bibitem[Stanimirovi\'c(2002)]{s02} Stanimirovi\'c, S. 2002, Short-Spacings Correction from the Single-Dish Perspective, ASPC, 278, 375, eds. Stanimirovi\'c et al.


\bibitem[Tsuboi et al.(1986)]{t86} Tsuboi, M., et al. 1986, AJ, 92, 818

\bibitem[Tsuboi et al.(1995)]{t95} Tsuboi, M., Kawabata, T., Kasuga, T., Handa, T., Kato, T. 1995, PASJ, 47, 829

\bibitem[Uchida et al.(1994)]{u94} Uchida, K. I., et al.\ 1994, ApJ, 421, 505

\bibitem[Uyaniker et al.(2003)]{u03} Uyaniker, B., Landecker, T. L., Gray, A. D., \& Kothes, R. 2003, ApJ, 585, 785

\bibitem[van Langevelde et al.(1992)]{v92} van Langevelde, H. J., Frail, D. A., Cordes, J. M., \& Diamond, P. J. 1992, ApJ, 396, 686

\bibitem[Yusef-Zadeh(1986)]{y86} Yusef-Zadeh, F. 1986, Ph.D thesis, Columbia University

\bibitem[Yusef-Zadeh(2003)]{y03} ------ 2003, ApJ, 598, 325

\bibitem[Yusef-Zadeh \& Morris(1988)]{y88} Yusef-Zadeh, F. \& Morris, M. 1988, ApJ, 329, 729

\bibitem[Yusef-Zadeh et al.(1984)]{y84} Yusef-Zadeh, F., Morris, M., \& Chance, D. 1984, Nature, 310, 557

\bibitem[Yusef-Zadeh et al.(2004)]{y04} Yusef-Zadeh, F., Hewitt, J., \& Cotton, W.  2004, ApJS, 155, 421

\bibitem[Yusef-Zadeh et al.(1997)]{y97} Yusef-Zadeh, F., Wardle, M., \& Parastaran, P. 1997, ApJ, 475L, 119

\bibitem[Yusef-Zadeh et al.(2006)]{y05} Yusef-Zadeh, F., Wardle, M., Muno, M., Law, C., \& Pound, M. 2005, AdSpR, 35, 1074

\end{thebibliography}
\end{document}